\documentclass[aps,reprint,nofootinbib,nobibnotes,notitlepage,superscriptaddress,onecolumn,prd,
 amsmath,amssymb
]{revtex4-2}

\usepackage[caption=false]{subfig}
\usepackage{braket}
\usepackage{float}
\usepackage{lipsum}
\usepackage{graphicx}
\usepackage{dcolumn}
\usepackage{bm}
\usepackage{natbib}
\usepackage{hyperref}
\hypersetup{
	colorlinks = true,
    linkcolor = Red,
    urlcolor  = Red,
    citecolor = Red
}

\usepackage{microtype}

\usepackage{verbatim}
\usepackage[amssymb]{SIunits}
\usepackage{tabularx}
\usepackage[dvipsnames]{xcolor}
\usepackage{wasysym}

\usepackage{tikz}

\def\be{\begin{equation}}
\def\ee{\end{equation}}

\usepackage{color}
\definecolor{darkgreen}{RGB}{0,120,0}

\definecolor{darkgreen}{RGB}{0,120,0}
\newcommand{\resub}[1]{{#1}}
\usepackage{xspace}
\newcommand*{\polybin}{\selectfont\textsc{PolyBin3D}\xspace}

\newcommand{\Mpch}{h^{-1}\mathrm{Mpc}}
\newcommand{\hMpc}{h\,\mathrm{Mpc}^{-1}}

\newcommand{\delD}[1]{(2\pi)^3\delta_\mathrm{D}\left({#1}\right)}

\newcommand{\av}[1]{\left\langle{#1}\right\rangle} 

\newcommand{\vk}{\vec k}
\newcommand{\hk}{\hat{\vec k}}

\newcommand{\vp}{\vec p}
\newcommand{\vz}{\vec z}

\newcommand{\vs}{\vec s}

\newcommand{\vx}{\vec x}
\newcommand{\vw}{\vec w}
\newcommand{\hx}{\hat{\vec x}}
\newcommand{\vy}{\vec y}

\newcommand{\C}{\mathsf{C}}
\newcommand{\Si}{\mathsf{S}^{-1}}
\newcommand{\Sit}{\mathsf{S}^{-\dag}}
\newcommand{\F}{\mathcal{F}}
\newcommand{\G}{\mathcal{G}}

\newcommand{\Q}{\mathsf{Q}}

\newcommand{\N}{\mathsf{N}}
\newcommand{\Ci}{\mathsf{C}^{-1}}
\newcommand{\Ai}{\mathsf{A}^{-1}}
\newcommand{\A}{\mathsf{A}}
\newcommand{\tCi}{\tilde{\mathsf{C}}^{-1}}

\newcommand{\hy}{\hat{\vec y}}
\newcommand{\hn}{\hat{\vec n}}

\newcommand{\hs}{\hat{\vec s}}

\renewcommand{\vr}{\vec r}

\renewcommand{\P}{\mathsf{P}}

\def\beq{\begin{eqnarray}}
\def\eeq{\end{eqnarray}}
\let\vec\mathbf

\usepackage{empheq}
\newcommand*\widefbox[1]{\fbox{\hspace{2em}#1\hspace{2em}}}
\newcommand*\widefboxred[1]{\textcolor{red}{\fbox{\hspace{2em}#1\hspace{2em}}}}

\usepackage{listings}

\definecolor{dkgreen}{rgb}{0,0.6,0}
\definecolor{gray}{rgb}{0.5,0.5,0.5}
\definecolor{mauve}{rgb}{0.58,0,0.82}

\usepackage{empheq}

\def\mnras{\rm{MNRAS}}
\def\aj{\rm{AJ}}
\def\aap{\rm{A\&A}}

\def\apj{\rm{ApJ}}                 
\def\apjl{\rm{ApJ}}                
\def\apjs{\rm{ApJS}}               
\def\jcap{\rm{JCAP}}
\def\pasj{\rm{PASJ}}

\begin{document}

\title{{\Large PolyBin3D}:\\
{\large A Suite of Optimal and Efficient Power Spectrum and Bispectrum Estimators for Large-Scale Structure}}

\author{Oliver~H.\,E.~Philcox}
\email{ohep2@cantab.ac.uk}
\affiliation{Simons Society of Fellows, Simons Foundation, New York, NY 10010, USA}
\affiliation{Center for Theoretical Physics, Columbia University, New York, NY 10027, USA}
\affiliation{Department of Physics,
Stanford University, Stanford, CA 94305, USA}

\author{Thomas Fl\"{o}ss}
\email{tsfloss@gmail.com}
\affiliation{Department of Mathematics, University of Vienna, 
1090 Vienna, Austria}
\affiliation{Department of Astrophysics, University of Vienna, 
1180 Vienna,
Austria}
\affiliation{Van Swinderen Institute, University of Groningen, 
9747 AG Groningen, The Netherlands}
\affiliation{Kapteyn Astronomical Institute, University of Groningen, 
9700 AV Groningen, The Netherlands}

\begin{abstract} 
    \noindent
    By measuring, modeling and interpreting cosmological datasets, one can place strong constraints on models of the Universe. Central to this effort are summary statistics such as power spectra and bispectra, which condense the high-dimensional data into low-dimensional representations. In this work, we introduce a modern set of estimators for computing such statistics from three-dimensional clustering data, and provide a flexible \textsc{Python/Cython} implementation; \href{https://github.com/oliverphilcox/PolyBin3D}{\polybin}. Working in a maximum-likelihood formalism, we derive general estimators for the two- and three-point functions, which yield unbiased spectra regardless of the survey mask and weighting scheme. These can be directly compared to theory without the need for mask-convolution. Furthermore, we present a numerical scheme for computing the optimal (minimum-variance) estimators for a given survey, which is shown to reduce error-bars on large-scales. Our \textsc{Python} package includes both general ``unwindowed'' estimators and their idealized equivalents (appropriate for simulations), each of which are efficiently implemented using fast Fourier transforms and Monte Carlo summation tricks, and additionally supports GPU acceleration \resub{using \textsc{jax}}. These are extensively validated in this work, with Monte Carlo convergence (relevant for masked data) achieved using only a small number of iterations (typically $<10$ for bispectra). This will allow for fast and unified measurement of two- and three-point functions from current and upcoming survey data. 
\end{abstract}

\maketitle

\section{Introduction}
\noindent From the clustering of galaxies to the distribution of gravitational waves, random fields play a major role in cosmology. Usually, one is concerned not with the precise stochastic realization but its ergodic distribution: central to this effort is the set of $N$-point \textit{correlation functions} or their Fourier-conjugates, \textit{polyspectra}, which characterize the distribution's moments. \resub{If the random field} in question follows a close-to-Gaussian distribution, the statistical properties of the field can be well-described by the set of correlators with low (integer) $N$, with a perfectly Gaussian distribution requiring only $N=1,2$. Three canonical examples of this are the temperature fluctuations probed by cosmic microwave background (CMB) experiments \citep[e.g.,][]{2020A&A...641A...6P,ACT:2020gnv,SPT-3G:2022hvq}, the distribution of matter sourcing weak gravitational lensing \citep{DES:2021wwk,Hildebrandt:2016iqg,Dalal:2023olq}, and the large-scale structure (LSS) observed in spectroscopic galaxy surveys \citep[e.g.,][]{2017MNRAS.470.2617A,2021PhRvD.103h3533A}. For the latter two observables, this limit is realized only on large-scales; on smaller scales, non-linear phenomena abound, and the distribution becomes highly non-Gaussian.

The above discussion motivates a practical manner in which to perform cosmological analyses: one measures the correlation functions of a given field and compares them to theoretical predictions. Indeed, this is the approach adopted by almost all analyses of CMB and LSS data to date \citep[e.g.,][]{2020A&A...641A...6P,ACT:2020gnv,SPT-3G:2022hvq,DES:2021wwk,Hildebrandt:2016iqg,Dalal:2023olq,2017MNRAS.470.2617A,2021PhRvD.103h3533A,Philcox:2021kcw}. Given a sufficiently accurate model, one can constrain any ingredient of the cosmological model used to generate it; in practice, this has allowed constraints to be wrought on the various components of the standard $\Lambda$CDM model \citep[e.g.,][]{2020A&A...641A...6P,ACT:2020gnv,SPT-3G:2022hvq,Philcox:2021kcw,Ivanov:2023qzb,2020JCAP...05..005D,Chen:2022jzq,Chen:2021wdi}, the physics of inflation \citep[e.g.,][]{Planck:2019kim,Cabass:2022wjy,Cabass:2022ymb,DAmico:2022gki,Chen:2024bdg}, and novel phenomena stemming from a variety of sources \citep[e.g.,][]{Rogers:2023ezo,Philcox:2022hkh,Philcox:2023ypl,Cabass:2022oap,Ivanov:2020ril,He:2023oke}. Many analyses consider only the two-point function (or power spectrum); however, the addition of the next-order statistics (the Fourier-space bispectrum and trispectrum, with $N=3,4$) have been shown to source tighter constraints on the \resub{$\Lambda$CDM} universe (in LSS analyses), as well as to directly the dynamics and field content of the primordial Universe \citep[e.g.,][]{Arkani-Hamed:2015bza,Liu:2019fag,Cheung:2007st}. In the non-linear regime probed by small-scale LSS experiments, the low-order correlators do not form a (close-to) complete basis; whilst there is little consensus about the optimal statistics in this regime, the literature abounds with possibilities, including one-point functions \citep[e.g.,][]{Chudaykin:2022sdl,Uhlemann:2019gni,Boyle:2020bqn}, void statistics \citep[e.g.,][]{Sheth:2003py,Pisani:2019cvo,Thiele:2023oqf}, marked correlators \citep[e.g.,][]{2005MNRAS.364..796S,2016JCAP...11..057W,2021PhRvL.126a1301M,2020PhRvD.102d3516P}, the power spectra of transformed fields \citep[e.g.,][]{1992MNRAS.254..315W,2009ApJ...698L..90N,2011ApJ...735...32W,2021JCAP...03..070R,Eisenstein:2006nk}, topological descriptors \citep[e.g.,][]{1987ApJ...319....1G,Matsubara:1994wn,1996cceu.conf...45M,Schmalzing:1997cv,SDSS:2003xnk,2014ApJ...796...86P,2022arXiv220308262B,Me94,Me94b,Sousbie:2007pn}, wavelets \citep[e.g.,][]{Valogiannis:2023mxf,Eickenberg:2022qvy,Blancard:2023iab,Pedersen:2023eop}, convolutional neural networks \citep[e.g.,][]{Giusarma:2019feb,Sharma:2024pth,Lemos:2023myd, Floss:2023ylq}, and beyond.

A central part of the aforementioned analyses is the measurement of $N$-point correlators from data. In principle, this is straightforward: one simply correlates the values of the random field at $N$ points in space, averaging over translations. In practice, many subtleties arise. What weights should be applied to the data before computing moments? Can one account for systematic effects and survey geometry? How can the estimator be efficiently implemented? In short, one seeks the optimal estimator for a given statistic, or at least, some close-to-optimal form that can be computed within reasonable computation time. 

In this work, we focus on the (quasi-)optimal estimation of the lowest-order Fourier-space correlators; the (binned) power spectrum and bispectrum. Furthermore, we will focus on three-dimensional scalar observables, such as the observed or simulated distribution of dark matter and galaxies; \citep{Philcox:2023uwe,Philcox:2023psd} (themselves building on \citep{Bucher:2015ura,2011MNRAS.417....2S,2015arXiv150200635S}) details the analogous estimators for scalar and tensor fields on the two-sphere. Ours is far from the first work to consider such estimators; on the contrary, there exists a large body of literature discussing such methods across several decades. These include a variety of (quasi-)optimal quadratic power spectrum estimators \citep{1998ApJ...499..555T,Hamilton:2005kz,Hamilton:2005ma,Hamilton:1999uw} (which saw significant application in the early 2000s \citep[e.g.,][]{2000MNRAS.317L..23H,2002MNRAS.335..887T,2002ApJ...571..191T,2004ApJ...606..702T}), approximate power spectrum weighting schemes \citep{1994ApJ...426...23F,2003ApJ...595..577Y,2006PASJ...58...93Y,2015MNRAS.453L..11B,2017JCAP...07..002H,2017JCAP...04..029S}, sub-optimal but efficient bispectrum estimators \citep{2012PhRvD..86f3511F,2011arXiv1105.2791F,Hung:2019ygc,2014JCAP...05..048C,2015PhRvD..91d3530S,2017MNRAS.472.2436W,2020MNRAS.492.1214P,2013PhRvD..88f3512S,2015MNRAS.451..539G,2021JCAP...03..105B}, as well as extension to anisotropy \citep{2017JCAP...12..020R,2019MNRAS.484..364S,2020JCAP...06..041G,2021arXiv210403976G,Byun:2022rvn,Gagrani:2016rfy,Rizzo:2022lmh,Yankelevich:2018uaz,2020MNRAS.497.1684S}, and the analogous methods for two-sphere observables (including \citep{1997PhRvD..55.5895T,1997ApJ...480...22T,1998PhRvD..57.2117B,1999PhRvD..59b7302B,1999ApJ...510..551O} for the two-point function and \citep{Fergusson:2010gn,Regan:2010cn,Sohn:2023fte,2011MNRAS.417....2S,2015arXiv150200635S,Shiraishi:2014roa,2003MNRAS.341..623S,Philcox4pt1,Philcox4pt2} for higher-point functions). This work adds to the canon in the following manners:
\begin{itemize}
	\item \textbf{Mask-Induced Bias}: Following methods developed for two-dimensional CMB analyses \citep{1998ApJ...499..555T,Hamilton:2005kz,Hamilton:2005ma,Hamilton:1999uw,Philcox:2023uwe,Philcox:2023psd,2011MNRAS.417....2S,2015arXiv150200635S,Philcox4pt1}, we develop \textit{unbiased} estimators for the power spectrum and bispectrum. This stands in contrast to most conventional approaches, whose outputs are modulated by the observational window function (\textit{i.e.}\ mask), stemming from the galaxy selection, bright stars in the image, dust extinction from the Milky Way and beyond. The resulting \textit{unwindowed} estimators can be efficiently computed, and allow the output spectra to be directly compared to theory. This stands in contrast to the standard approach \citep[e.g.,][]{1994ApJ...426...23F,2017JCAP...07..002H}, which computes mask-convolved spectra (pseudo-spectra), that must be compared to similarly convolved theory (see \citep{2004MNRAS.349..603E} for a detailed discussion of these differences). The latter approach is extremely expensive for statistics beyond the power spectrum, which has typically led to works making simplifying assumptions \citep[e.g.,][]{2015MNRAS.451..539G,2020JCAP...05..005D,2017MNRAS.465.1757G,DAmico:2022gki} (though see \citep{Pardede:2022udo} for an improved approach), which can induce significant bias on cosmological parameters, recently demonstrated in \citep{Chen:2024bdg}.
	\item \textbf{Weighting \& Optimality}: Our estimators allow the data to be weighted by arbitrary (linear) schemes, whilst remaining unbiased. This \resub{could facilitate} anisotropic noise weighting, deprojection of systematics \citep[e.g.,][]{Alonso:2018jzx,Bahr-Kalus:2018rom}), Wiener-filtering, \resub{the filling of small-holes in the map via in-painting to avoid Fourier-space ringing \citep[e.g.,][]{Gimeno-Amo:2024hca}} and beyond to be applied to the data. Furthermore, we do not require an explicit form for the weighting, only its action on a map. Finally, we provide a numerical approach for computing the optimal weighting scheme (which may be dense in \resub{both} real- and Fourier-space), which gives the minimum-variance estimator (in the Gaussian limit), and thus the tightest error-bars on any derived parameters.
	\item \textbf{Holes}: We pay close attention to holes in the mask, \textit{i.e.}\ regions with vanishing background density. These can lead to biases and instabilities in the optimal estimators (found in previous treatments \citep{2021PhRvD.103j3504P,Philcox:2021ukg}) unless carefully accounted for.
	\item \textbf{Generalization}: We carefully consider the simplification of our estimators in limiting regimes, such that they can be efficiently applied to both observational data and numerical simulations (scaling as \resub{$\mathcal{O}(N_{\rm pix}\log N_{\rm pix})$} when using $N_{\rm pix}$ pixels). We also simultaneously consider all the main statistics used in modern full-shape analyses of galaxy survey data \citep[e.g.,][]{Philcox:2021kcw,Ivanov:2023qzb}: the anisotropic moments of both the power spectrum and bispectrum. The corresponding forms could be straightforwardly extended to the trispectrum. 
	\item \textbf{Code}: We provide a modular and easy-to-use \textsc{Python} package, \polybin, implementing the unwindowed power spectrum and bispectrum estimators, as well as their idealized equivalents. This makes extensive use of fast Fourier transforms, \resub{parallelized \textsc{cython} code} and Monte Carlo methods, which allow the high-dimensional equations to be computed in a very efficient manner. \resub{We additionally provide
    GPU support using JAX, which can significantly enhance performance, with only a minimal addition of code.} 
    We provide a suite of tutorials\footnote{Available at \href{https://github.com/oliverphilcox/PolyBin3D}{GitHub.com/OliverPhilcox/PolyBin3D}.} demonstrating and validating our pipeline.
\end{itemize}
Finally, we note that this work builds upon our previous formulations of optimal power spectrum and bispectrum estimators \citep{Philcox:2023uwe,Philcox:2023psd} (for the CMB) and \citep{2021PhRvD.103j3504P,Philcox:2021ukg,Ivanov:2023qzb} (for galaxy clustering). Our treatment here is purposefully self-contained, and extends beyond the former in many ways, including (but not limited to): generalized treatment of weighting schemes, holes in the survey mask, improved Monte Carlo methods, a unified approach to all statistics, extensive validation tests, and, last but certainly not least, a new user-friendly CPU and GPU code, \polybin.

\vskip 4pt
The remainder of this paper is organized as follows. In \S\ref{sec: estimator-theory} we describe the theoretical underpinnings of the (quasi-)optimal estimators, before discussing the specific application to LSS power spectrum and bispectrum estimators in \S\ref{sec: power-spectra}\,\&\,\S\ref{sec: bispectra}. \S\ref{sec: implementation} discusses our numerical implementations in \polybin, before the estimators are validated in \S\ref{sec: validation}. We conclude with a general discussion in \S\ref{sec: conclusion}. Throughout this work we will define the forward and reverse Fourier transforms by
\beq
    a(\vk) \equiv \mathrm{FT}[a](\vk) = \int {\rm d}\vx\,e^{-i\vk\cdot\vx}a(\vx), \qquad a(\vx) \equiv \mathrm{IFT}[a](\vx)= \int_{\vk}e^{i\vk\cdot\vx}a(\vk),
\eeq
indicating configuration- and Fourier-space quantities by the coordinates $\vx,\vy,\vr,\cdots$ and $\vk,\vp,\cdots$ respectively. Here and henceforth we notate $\int_{\vk}\equiv \int {\rm d}\vk/(2\pi)^3$. Key equations throughout the text are boxed, with those defining the key parts of the \polybin unwindowed and ideal estimators shown in \textcolor{red}{red} and \textcolor{blue}{blue} respectively.

\section{Quasi-Optimal Polyspectrum Estimators}\label{sec: estimator-theory}
\noindent We begin by discussing the general form of quasi-optimal estimators for binned polyspectra. Much of our treatment follows \citep{Philcox:2023uwe,Philcox:2023psd} (itself building on \citep{2021PhRvD.103j3504P,Philcox:2021ukg,Ivanov:2023qzb,Hamilton:1999uw,Hamilton:2005kz,Hamilton:2005ma}), but we recapitulate it for clarity. Furthermore, we introduce a more \resub{optimal} weighting scheme and pixelation treatment, and take special care to ensure that our estimators are stable in the presence of holes in the survey mask. In particular, previous treatments required dividing by the survey mask (as well as recursive application of pixellation matrices), which can cause numerical problems when the mask is incomplete, and induce non-trivial bias; our new approach obviates these issues and is guaranteed to be unbiased. Whilst this section will be technical in nature (and could be skipped by the reader uninterested in theoretical underpinnings), it sets the form of the estimators used in the remainder of this work.

\subsection{Optimal Estimators}\label{subsec: optimal-estimators}
\noindent At heart, estimator theory seeks to answer the following question: ``how can I obtain the minimum-variance unbiased estimator for given set of quantities, $\{x_\alpha\}$, from a dataset, $d$?''. To answer this, we must understand the statistical properties of $d$, and their relation to the quantities of interest, $\{x_\alpha\}$ (such as power spectrum bandpowers). In this work, our dataset is the large-scale structure density field extracted from a set of simulations or data, which we will denote $d_i$ for pixel index $i$. On sufficiently large-scales this is fully described by a Gaussian likelihood with covariance $\C_{ij}\equiv\av{d_id_j^*}$; on smaller (but still perturbative) scales, $d$ follows an Edgeworth expansion \citep[e.g.,][]{2017arXiv170903452S} with log-likelihood (assuming Einstein summation):
\beq\label{eq: edgeworth}
    \log\mathcal{L}(d) &=& \left(-\frac{1}{2}\Ci_{ij}d^{i*}d^{j}-\frac{1}{2}\mathrm{Tr}\log \C\right)\\\nonumber
    &&\,+\,\log\left[1+\frac{1}{3!}\left(\mathcal{H}^{ijk}[\Ci d]\right)^*\av{d_id_jd_k}_c+\frac{1}{4!}\left(\mathcal{H}^{ijkl}[\Ci d]\right)^*\av{d_id_jd_kd_l}_c+\cdots\right]+\text{const.},
\eeq
which is specified entirely by the connected correlation functions $\C_{ij}$ and $\av{\cdots}_c$.  Here, we have defined the Hermite tensors:
\beq
    \mathcal{H}_{ijk}[h] &=& h_ih_jh_k - \left[\av{h_ih_j}h_k+\text{2 perms.}\right]\\\nonumber
    \mathcal{H}_{ijkl}[h] &=& h_ih_jh_kh_l - \left[\av{h_ih_j}h_kh_l+\text{5 perms.}\right]+\left[\av{h_ih_j}\av{h_kh_l}+\text{2 perms.}\right],
\eeq
and note that only the first line in \eqref{eq: edgeworth} survives in the Gaussian limit.

According to the Cram\'{e}r-Rao theorem, the optimal estimator for some quantity $x_\alpha$ appearing only in the $n$-point correlation function (such as a coefficient in the binned $n$-point function) can be obtained by extremizing \eqref{eq: edgeworth}. Assuming that the background solution is Gaussian (\textit{i.e.}\ that the fiducial correlators are zero for $n>2$), this gives the estimator
\begin{empheq}[box=\widefbox]{align}\label{eq: opt-estimator}
    \hat{x}_\alpha = \sum_\beta\F_{\alpha\beta}^{-1}x^{\rm num}_{\beta}, \qquad x_\alpha^{\rm num} = \frac{1}{n!}\frac{\partial\av{d^{i_1}\cdots d^{i_n}}_c}{\partial x_\alpha}\mathcal{H}^*_{i_1\cdots i_n}[\Ci d]\\\nonumber
    \F_{\alpha\beta} = \frac{1}{n!}\left[\frac{\partial\av{d^{i_1}\cdots d^{i_n}}^*_c}{\partial x_\alpha}\Ci_{i_1j_1}\cdots \Ci_{i_nj_n}\frac{\partial\av{d^{j_1}\cdots d^{j_n}}_c}{\partial x_\beta}\right]^*
\end{empheq} 
in terms of a numerator, $x^{\rm num}$ and a `Fisher matrix' $\F_{\alpha\beta}$.\footnote{For $n=2$, these estimators are formed by expanding around some fiducial $\{x_\alpha\}$ rather than $\{x_\alpha=0\}$, \textit{i.e.}\ computing the Newton-Raphson estimate \citep[cf.,][]{2004MNRAS.349..603E,Hamilton:1999uw,Hamilton:2005kz,Hamilton:2005ma}. This gives an extra term in the estimators, but it cancels with that from the trace in \eqref{eq: edgeworth}, such that \eqref{eq: opt-estimator} applies for all $n\geq 2$.} Note that we have allowed the data to be complex to retain generality, though $d_i$ is usually real in LSS contexts. \eqref{eq: opt-estimator} obeys certain properties: 
\begin{itemize}
    \item \textbf{Bias}: If $\{x_\alpha\}$ fully describe the $n$-point correlator (such that $\av{d_{i_1}\cdots d_{i_n}} = \sum_\alpha x_\alpha\,\partial\av{d_{i_1}\cdots d_{i_n}}/\partial x_\alpha $, the estimator is unbiased, such that $\mathbb{E}[\hat{x}_\alpha] = x_\alpha$.
    \item \textbf{Optimality}: Under the assumption of Gaussianity, \eqref{eq: opt-estimator} is the minimum-variance estimator for $x_\alpha$, and thus optimal.
    \item \textbf{Covariance}: Again working in the Gaussian limit, the covariance of $\hat{x}_\alpha$ is given by $\left[\F^{-1}\right]_{\alpha\beta}$.
\end{itemize}
In practice, implementing the optimal estimator of \eqref{eq: opt-estimator} is non-trivial since (a) there are often contributions not described by $\{x_\alpha\}$ (e.g., stochasticity), and (b) the necessary operations are too high-dimensional to implement in a na\"ive fashion. For example, the pixel-space covariance $\C_{ij}$ has dimension $N_{\rm pix}\times N_{\rm pix}$ and $N_{\rm pix}\gtrsim 10^6$ usually, making it infeasible to store, let alone invert. In the remainder of this paper, we discuss how such issues can be overcome, leading to practically implementable and \resub{close-to-optimal} estimators.

\subsection{Generalized Estimators}\label{subsec: gen-estimators}
\noindent Starting from \eqref{eq: opt-estimator}, a more general estimator can be wrought by replacing the $\Ci$ weighting (which, in general, is computationally limiting) with a general matrix $\tCi$ (which may be asymmetric). For example, one may choose $\tCi$ to be diagonal in configuration- or Fourier-space, which would greatly reduce its computational requirements. \resub{Furthermore, it is useful to introduce the \textit{pointing matrix}, $\mathsf{P}$, which encodes the response of the dataset $d$ to the observational quantity of interest, $\delta$ (e.g., the fractional overdensity field):
\beq\label{eq: pointing}
    d^i = \mathsf{P}^{ij}\delta_j + \epsilon^i,
\eeq
where $\epsilon^i$ is some noise contribution. Inserting \eqref{eq: pointing} into \eqref{eq: opt-estimator} and replacing $\Ci$ with the generalized weight, we note that $\tCi$ only enters multiplied by $\mathsf{P}^\dagger$: this motivates the introduction of a new weighting matrix, $\Si \equiv \mathsf{P}^\dagger\tCi$. With these modifications, \eqref{eq: opt-estimator} becomes
\begin{empheq}[box=\widefbox]{align}\label{eq: gen-estimator-Sinv}
    \qquad\qquad\hat{x}_\alpha &\equiv \sum_\beta\F_{\alpha\beta}^{-1}\left[x^{\rm num}_{\beta}-x^{\rm bias}_{\beta}\right]\\\nonumber
    x_\alpha^{\rm num} &=\frac{1}{n!}\frac{\partial\zeta^{i_1\cdots i_n}}{\partial x_\alpha}\mathcal{H}^*_{i_1\cdots i_n}[\Si d],\\\nonumber
    x_\alpha^{\rm bias} &= \frac{1}{n!}\frac{\partial\zeta^{i_1\cdots i_n}}{\partial x_\alpha}\left[\Si_{i_1j_1}\cdots \Si_{i_nj_n}\N^{j_1\cdots j_n}\right]^*\\\nonumber
    \F_{\alpha\beta} &= \frac{1}{n!}\left[\frac{\partial\zeta^{i_1\cdots i_n,*}}{\partial x_\alpha}[\Si\mathsf{P}]_{i_1j_1}\cdots [\Si\mathsf{P}]_{i_nj_n}\frac{\partial\zeta^{j_1\cdots j_n}}{\partial x_\beta}\right]^*.
\end{empheq}
for primordial correlator $\zeta^{i_1\cdots i_n} \equiv \av{\delta^{i_1}\cdots\delta^{i_n}}_c$. Here}, we have additionally introduced a `bias' term $\N^{i_1\cdots i_n}$; this is defined as the piece of $\av{d^{i_1}\cdots d^{i_n}}_c$ not captured by $\{x_\alpha\}$ (usually a shot-noise contribution). Finally, we note that $\F_{\alpha\beta}$ is Hermitian \resub{only if $\Si\mathsf{P}$ is}. 

Our new estimators have the following properties:
\begin{itemize}
    \item \textbf{Bias}: The estimator is unbiased regardless of our assumptions on the likelihood and \resub{the weighting scheme $\Si$. This holds by construction due to our choice of normalization $\F$.}
    \item \textbf{Covariance}: In the Gaussian limit, the covariance of ${x}^{\rm num}_\alpha$ is given by \resub{
    \beq\label{eq: gen-cov}
        \mathrm{cov}\left(x^{\rm num}_\alpha, x^{\rm num}_\beta\right) = \frac{1}{n!}\left[\frac{\partial\av{\delta^{i_1}\cdots \delta^{i_n}}^*_c}{\partial x_\alpha}[\Si\C\Sit]_{i_1j_1}\cdots [\Si\C\Sit]_{i_nj_n}\frac{\partial\av{\delta^{j_1}\cdots \delta^{j_n}}_c}{\partial x_\beta}\right]^*
    \eeq
    (with \resub{$\mathsf{X}^{-\dag} \equiv (\mathsf{X}^{-1})^\dag$} here and henceforth). This takes a similar form to the normalization matrix and can be estimated similarly.}
    \item \textbf{Optimality}: \resub{In the limit of
    \beq\label{eq: general-Sinv-optimality}
        \Si\C\Sit=\Si\mathsf{P},
    \eeq
    the covariance of $\widehat{x}_\alpha$ is equal to the inverse normalization, $\F^{-1}$.} Assuming a Gaussian likelihood, this implies that \eqref{eq: gen-estimator-Sinv} is the minimum-variance estimator \resub{for $\{x_\alpha\}$, since its saturates the Cram\'er-Rao bound. If $\C$ is invertible, this requires $\Si= \mathsf{P}^\dagger \Ci$.}
\end{itemize}

Though it may not seem immediately apparent, these estimators are far simpler to implement than the general forms of \eqref{eq: opt-estimator} since there is no requirement to invert $\C$; furthermore, the numerators involve only a single transformation applied to the data: $d\to \Si d$. These estimators will be used throughout the remainder of this work, and represent an important difference to previous prescriptions \citep{2021PhRvD.103j3504P,Philcox:2021ukg}, which required explicit (and unstable) division by the window function, as well as repeated convolutions \resub{with the pixel window function}.

\subsection{Specialization to Spectroscopic Surveys}\label{subsec: spec-estimators}
\noindent In the above sections, we have presented formal estimators for quantities appearing in the $n$-point correlators of a generic field $d_i$. Here, we will consider their application to three-dimensional surveys, such as the galaxy density measured by spectroscopic surveys or the matter density field from simulations. Whilst we leave the details of the precise $n$-point estimators to \S\ref{sec: power-spectra}\,\&\,\S\ref{sec: bispectra}, we will here discuss the form of the data, \resub{the pointing matrix $\mathsf{P}$ and the covariance $\C$}, since these impact estimators of all order. Throughout this section we will \textit{not} assume that $\C$ is \resub{invertible}, unlike previous works. This is discussed in more detail in \S\ref{subsec: weighting}. 

In spectroscopic surveys, the observational data, $d$, is taken as the difference between a set of observed galaxies, $n_g$, and random points, $n_r$. In configuration-space, this can be written
\beq\label{eq: data-model}
    d(\vx) &=& \int {\rm d}\vy\,m(\vx-\vy)\left[n_g(\vy)-\alpha\,n_r(\vy)\right] = \resub{\sum_{i=1}^{N_g}w_{g,i}m(\vx-\vx_{g,i}) - \alpha\sum_{i=1}^{N_r}w_{r,i}m(\vx-\vx_{r,i})}\\\nonumber
    &\equiv& \resub{\int {\rm d}\vy\,m(\vx-\vy)n(\vy)\delta(\vy)+\epsilon(\vx)}
\eeq
\resub{expressing $n_g$ as a sum over $N_g$ galaxy positions, $\{\vx_{g,i}\}$ with weights $w_g$ (which may include FKP weights), with an analogous form for $n_r$.}\footnote{\resub{Often, we work with pixel-deconvolved data; in this case $m(\vx-\vy) = \delta_{\rm D}(\vx-\vy)$. This limit also applies for continuous fields, such as simulation outputs.}} Here, $m(\vx)$ is a pixelation window function (e.g., cloud-in-cell), which encode how discrete particles are assigned to a lattice \citep[e.g.,][]{2005ApJ...620..559J} and $\alpha\equiv\sum_{i=1}^{N_g}w_{g,i}/\sum_{i=1}^{N_r}w_{r,i}$ is the ratio of galaxies to randoms. In the bottom line, we have converted these fields into the underlying overdensity, $\delta$ (which is the cosmological quantity of interest), a Poissonian stochasticity field $\epsilon$,\footnote{We absorb any non-Poissonian noise into the overdensity field $\delta$.} and the \resub{(weighted)} background galaxy density $n$, assuming that $\av{\delta}=\av{\epsilon}=0$. 

\resub{From \eqref{eq: data-model}, we can extract the pointing matrix:
\beq\label{eq: pointing}
    \mathsf{P}(\vx,\vy) = m(\vx-\vy)n(\vy)
\eeq
which simply multiplies the overdensity by the background distribution of galaxies (which may contain holes) and convolves with the pixelation mask. We note that the normalization matrix of \eqref{eq: gen-estimator-Sinv}, $\F_{\alpha\beta}$,} includes $n$ factors of $\P$ and thus the mask $n(\vx)$, encoding the effective volume of the statistic. Our data model applies also to simulations: in this case, the background density is uniform ($n(\vx) = \bar{n}$), and, for the matter field, $\epsilon(\vx)$ can usually be set to zero. 

The optimal estimators discussed in \S\ref{subsec: optimal-estimators} require the two-point correlation function of the data $\C\equiv \av{dd^\dag}$. This is given in matrix form by
\beq\label{eq: cov-def}
    \resub{\mathsf{C} = \P\xi\P^\dag+\mathsf{N}},
\eeq
where $\xi = \av{\delta\delta^\dag}$ is the true two-point function of the tracer (encoding the power spectrum) and $\mathsf{N}\equiv\av{\epsilon\epsilon^\dagger}$ is a Poisson noise term, which \resub{depends on the mask and pixelation window. For spectroscopic surveys, the noise can be written
\beq\label{eq: two-point-noise}
    \mathsf{N}(\vx,\vy) &=& \left(1+\alpha^2/\alpha_2\right)\int{\rm d}\vz\,m(\vx-\vz)m(\vy-\vz)\overline{n}_2(\vz), \qquad \alpha_2 = \sum_{i=1}^{N_g}w_{g,i}^2 \big/\sum_{i=1}^{N_r}w_{r,i}^2
\eeq
where $\overline{n}_2$ is the background density corresponding to a doubly-weighted galaxy sample, and the prefactor accounts for the relative number of galaxies and random points in the dataset. In practice, this can be estimated using a catalog of doubly-weighted random points:
\beq
    n_2(\vx) &\equiv& (\alpha_2+\alpha^2)\sum_{i=1}^{N_r}w_{r,i}^2\delta_{\rm D}(\vx-\vx_{r,i}),
\eeq
with expectation $\av{n_2(\vx)} = (1+\alpha^2/\alpha_2)\overline{n}(\vx)$.\footnote{\resub{This is simply derived by considering the contributions to $d(\vx)d(\vy)$ from coincident galaxies and random points and writing the result in terms of the random catalog}.} We note that $\N(\vx,\vy)\propto \delta_{\rm D}(\vx-\vy)$ in the absence of a pixel window function.}


\subsection{Weighting Schemes}\label{subsec: weighting}
\noindent As discussed in \S\ref{subsec: gen-estimators}, optimal polyspectrum estimators involve the weighting scheme \resub{$\Si_{\rm opt} = \mathsf{P}^\dagger\Ci$}, which requires inverting \eqref{eq: cov-def}. \resub{In some settings, including the analysis of uniform-density simulations, this can be computed analytically, whilst in others we must resort to simple approximations or numerical tricks. Below, we consider three cases: ideal weights (relevant to simulations), FKP weights (appropriate for some spectroscopic scenarios), and optimal weights (which give minimum-variance estimators). We remind the reader that our polyspectrum estimators are unbiased for \textit{any} weighting; however, a poor choice of $\Si$ will lead to high-noise measurements.}

\vskip 8pt
\paragraph{Ideal Weights} If the background density is constant, $n(\vx) = \bar{n}$, we can invert the covariance matrix of \eqref{eq: cov-def} directly:
\beq
    \Ci \to \frac{1}{\bar{n}^2}\mathsf{M}^{-\dag}[\xi +\mathsf{I}/\bar{n}]^{-1}\mathsf{M}^{-1},
\eeq
where $\mathsf{I}$ is the identity \resub{and $\mathsf{M}$ represents the (invertible) pixel convolution operation}. This takes a straightforward form in Fourier-space:
\beq\label{eq: ideal-Ci}
    \Ci_{\rm ideal}(\vx,\vy) = \frac{1}{\bar{n}^2}\int_{\vk}e^{i\vk\cdot(\vx-\vy)}\frac{1}{P(\vk)+\bar{n}^{-1}}\frac{1}{m^2(\vk)},
\eeq
involving the fiducial power spectrum $P(\vk)+1/\bar{n}$. The associated minimum-variance filter is given by
\beq\label{eq: Sinv-ideal}
    \boxed{\Si_{\rm ideal}(\vx,\vy) = \frac{1}{\bar{n}}\int_{\vk}e^{i\vk\cdot(\vx-\vy)}\frac{1}{P(\vec k)+\bar{n}^{-1}}\frac{1}{m(\vk)}} \qquad \Rightarrow \qquad [\Si_{\rm ideal}v](\vx) = \frac{1}{\bar{n}}\int_{\vk}e^{i\vk\cdot\vx}\frac{1}{P(\vec k)+\bar{n}^{-1}}\frac{v(\vk)}{m(\vk)},
\eeq
where the second equality shows how $\Si$ can be applied to an arbitrary map $v(\vx)$; \resub{we Fourier transform $v$, divide by the pixel window and the fiducial power spectrum, then return to configuration-space and normalize appropriately.} This is closely related to a Wiener filter
\citep[e.g.,][]{Fergusson:2011sa,2012PhRvD..86f3511F} and will be used to build the idealized power spectrum and bispectrum estimators used in \polybin.

\vskip 8pt
\paragraph{FKP Weights}
A second common limit is the ``FKP'' form \citep{1994ApJ...426...23F}. This arises when one models the covariance assuming a constant power spectrum $P(\vec k)=P_{\rm FKP}$ but allowing for a varying background density, $n(\vx)$. \resub{Adopting this model, we can obtain a quasi-optimal filtering $\Si_{\rm FKP}\equiv \P^\dagger\C_{\rm FKP}^{-1}$ with}
\beq\label{eq: Sinv-fkp}
    \boxed{\Si_{\rm FKP}(\vx,\vy) = \frac{1}{n(\vx)P_{\rm FKP}+1}\int_{\vk}e^{i\vk\cdot(\vx-\vy)}\frac{1}{m(\vk)},}
\eeq
\resub{which deconvolves the mask then weights the data by $w_{\rm FKP}(\vx) 
\equiv [n(\vx)P_{\rm FKP}+1]^{-1}$.} This is similar to the usual FKP factor but includes both angular and redshift variations in the background density. \resub{Often, one applies the FKP weight to the dataset before it is assigned to a 3D grid (\textit{i.e.}\, it is absorbed within the $w_{g,r}$ factors in \ref{eq: data-model}); in this case, the FKP weight is absorbed into the background density $n(\vx)$, and the weighting becomes trivial:
\beq
    \Si_{\rm FKP}(\vx,\vy) = \int_{\vk}e^{i\vk\cdot(\vx-\vy)}\frac{1}{m(\vk)}.
\eeq} 

\vskip 8pt
\paragraph{Optimal Weights}
\resub{Beyond the above limits, the covariance matrix is difficult to invert, being dense in both configuration- and Fourier-space. Furthermore, the background tracer density $n(\vx)$ often contains holes, such that the matrix is not of full-rank and its inverse does not exist} \resub{(\textit{i.e.}\ $\C(\vx,\vy) = 0$ if $n(\vx)=0$ or $n(\vy)=0$, neglecting the pixel-window for simplicity).} This does not preclude its use however, since the estimators below require only its action on a map, $[\Si v](\vx)$, which can be computed using numerical techniques; moreover, the full optimality condition \eqref{eq: general-Sinv-optimality} can be satisfied even if $\C$ is not invertible.


To illustrate this, we first consider \eqref{eq: general-Sinv-optimality} in more depth.\footnote{\resub{An alternative approach is to rewrite the (sufficient) condition $\Si_{\rm opt} = \P^\dagger\Ci$ using the Woodbury matrix identity:
\beq
    \Si_{\rm opt} \equiv \P^\dagger\left(\P\xi\P^{\dagger}+\N\right)^{-1} &=& \xi^{-1}\left(\xi^{-1}+\P^{\dagger}\N^{-1}\P\right)^{-1}\P^\dagger\N^{-1} = \left(\mathsf{I}+\P^{\dagger}\N^{-1}\P\xi\right)^{-1}\P^\dagger\N^{-1}
\eeq
\citep[e.g.,][]{Oh:1998sr}, defining some inverse noise matrix $\N^{-1}$, which can be set to zero outside the mask. This still involves a complicated inverse which must be computed using numerical methods (e.g., conjugate gradient descent); however, it is simpler than before since $(\mathsf{1}+\P^\dag\N^{-1}\P\xi)(\vx,\vy)$ is non-zero everywhere.}}
\resub{For this purpose it is useful to factor out the pixel-window matrix $\mathsf{M}$; this can be achieved by defining pixel-deconvolved quantities according to $\P = \mathsf{M}\tilde{\mathsf{P}}$, $\Si = \tilde{\mathsf{S}}^{-1}\mathsf{M}^{-1}$ and $\N = \mathsf{M}\tilde{\mathsf{N}}\mathsf{M}^\dagger$.\footnote{Equivalently, one can simply deconvolve the pixel-window from the data, and set $\mathsf{M}=\mathsf{I}$ thereafter.}} Inserting \eqref{eq: cov-def} into the transpose of \eqref{eq: general-Sinv-optimality}, we find
\beq\label{eq: opt-transpose1}
    \tilde{\mathsf{S}}^{-1}_{ik}[\tilde{\mathsf{P}}\xi \tilde{\mathsf{P}}^\dagger +\tilde{\mathsf{N}}]_{kl}\tilde{\mathsf{S}}^{-\dagger}_{lj} = \tilde{\mathsf{P}}^\dagger_{ik}\tilde{\mathsf{S}}^{-\dagger}_{kj},
\eeq
writing out matrix components explicitly. Next, we note that $\tilde{\mathsf{P}}(\vx,\vy)$ vanishes for all $\vx$ with $n(\vx)=0$ (since we have deconvolved the pixel-window); the same is true for both axes of the noise covariance $\tilde{\mathsf{N}}$. As such, we can restrict the summation in \eqref{eq: opt-transpose1} to pixels with $n(\vx)\neq 0$, which we denote by capital roman indices $I,J,\ldots\in\{1,N_{\rm sub}\}$:
\beq\label{eq: tmp-mat}
    \tilde{\mathsf{S}}^{-1}_{IK}[\tilde{\mathsf{P}}\xi\tilde{\mathsf{P}}^\dagger+\tilde{\mathsf{N}}]_{KL}\tilde{\mathsf{S}}^{-\dagger}_{LJ} = \tilde{\mathsf{P}}^\dagger_{IK}\tilde{\mathsf{S}}^{-\dagger}_{KJ}.
\eeq
\resub{Notably, each quantity in \eqref{eq: tmp-mat} is an $N_{\rm sub}\times N_{\rm sub}$ matrix in the subspace of non-zero pixels; restricting to this space}, we require $\tilde{\mathsf{S}}^{-1}_{IK}[\tilde{\mathsf{P}}\xi+\tilde{\mathsf{N}}\tilde{\mathsf{P}}^{-\dagger}]_{KJ}=\mathsf{I}_{IJ}$ (noting that all subspace matrices are full-rank and invertible). Rearranging and applying to an arbitrary vector $\tilde{v}_I$ in the subspace leads to\footnote{This differs subtly but importantly from the form suggested in \citep{2021PhRvD.103j3504P}, which required solving for $\Si = \P^\dagger \tilde{\mathsf{C}}^{-1}$ via $[\P\xi \P^\dag  + \N]\tCi v = v$. The two approaches coincide for invertible $\P$.} 
\beq
    [\tilde{\mathsf{P}}\xi+\tilde{\mathsf{N}}\tilde{\mathsf{P}}^{-\dagger}]_{IJ}\tilde{\mathsf{S}}^{-1}_{JK}\tilde{v}_K=\tilde{v}_I
\eeq
Finally, this can be rewritten in the full $N_{\rm pix}$-dimensional space, by promoting $\tilde{v}_I$ to a \resub{(mask-deconvolved)} map $\tilde{v}(\vx)$ with $\tilde{v}(\vx)=0$ for $n(\vx)=0$.\footnote{This restriction is justified since $\tilde{\mathsf{S}}^{-1}$ acts only on maps which have this property (\textit{e.g.}\ $\mathsf{M}^{-1}d$).} Inserting the usual assumptions for the pointing matrix and noise term (\ref{eq: pointing} \& \ref{eq: two-point-noise}), this gives the integral form 
\beq\label{eq: cgd-operation-eqn}
    \int {\rm d}\vy\left[n(\vx)\xi(\vx,\vy)+\frac{n_2(\vx)}{n(\vx)}\delta_{\rm D}(\vx-\vy)\right]\Phi(\vy)[\Si v](\vy)= [\mathsf{M}^{-1} v](\vx),
\eeq
where $\Phi(\vx)=0$ if $n(\vx)=0$ and $1$ else, \resub{involving the ratio of the doubly- and singly-weighted background density. We can similarly form an equation for the transpose filtering, $\Sit$, (which we will need below):}
\beq\label{eq: cgd-transpose-operation-eqn}
    \Phi(\vx)\int{\rm d\vy}\left[\xi(\vx,\vy)n(\vy)+\frac{n_2(\vx)}{n(\vx)}\delta_{\rm D}(\vx-\vy)\right][\Sit v](\vy)= [\mathsf{M}^{-1}v](\vx).
\eeq

The above equations have an important and practical consequence. Given a map $v(\vx)$, one can compute $[\Si v](\vx)$ \resub{and $[\Sit v](\vx)$} numerically using conjugate gradient descent schemes \citep[cf.][]{2011MNRAS.417....2S,2021PhRvD.103j3504P} (or with more advanced Machine Learning prescriptions \citep[e.g.,][]{Munchmeyer:2019kng}), given the approximate forms in \eqref{eq: Sinv-ideal}\,\&\,\eqref{eq: Sinv-fkp}. This can be done efficiently noting that, if $\xi$ is translation-invariant, the equation can be implemented using Fourier-transforms (as we discuss in \S\ref{sec: power-spectra}). 

\vskip 8pt
Finally, let us remark on the journey so far. Starting from the perturbative likelihood for our dataset $d$, we have derived optimal estimators for binned polyspectrum coefficients ($\{x_\alpha\}$) and their generalized unbiased equivalents for arbitrary weighting schemes. Specializing to spectroscopic surveys with incomplete window functions, $n(\vx)$, we have motivated a more nuanced set of estimators \eqref{eq: gen-estimator-Sinv}, with an associated optimality condition on the weighting $\Si$, and have discussed how this can be computed, either via approximations or with numerical methods. In the remainder of this work, we will apply the above formalism to the problem at hand: estimating the redshift-space power spectrum and bispectrum \resub{coefficients} from observational clustering data and simulations.

\section{Power Spectrum Estimators}\label{sec: power-spectra}
\subsection{Definitions}
\noindent Our first application of the quasi-optimal estimators is to the power spectrum of spectroscopic data. To form the estimators we must first define the quantities we wish to measure and their relation to the pixel-space correlators: here the power spectrum bandpowers and the two-point function of the underlying overdensity field $\xi_{ij}\equiv\av{\delta^{\,}_i\delta_j^*}$.

In configuration-space, the two point function can be expressed in terms of the power spectrum, $P$, via
\beq\label{eq: cf-pk-def}
    \xi(\vx,\vy) \equiv \frac{1}{2}\int_{\vk}e^{i\vk\cdot(\vx-\vy)}\left[P(\vk;\hx)+P(-\vk;\hy)\right],
\eeq
where $P(\vk;\hat{\vec d})$ is the power spectrum with anisotropy measured with respect to the line-of-sight $\hat{\vec d}$. Here, we have symmetrized over the two `end-point' definitions for $\hn$, as in the Yamamoto approximation \citep{2006PASJ...58...93Y}.\footnote{See \citep{2018MNRAS.476.4403C,2021PhRvD.103l3509P,Castorina:2018nlb,Castorina:2019hyr} for more nuanced anisotropy schemes using $\hn = \widehat{\vx+\vy}$ or $(\hx+\hy)/2$; the resulting bandpowers could also be computed with similar estimators to the below.} In the distant-observer limit $\hx\to\hn$ for a global line-of-sight $\hn$, thus $\xi(\vx,\vy)$ is translation-invariant and thus diagonal in Fourier-space. Expanding the angular dependence of $P(\hk;\hat{\vec d})$ in Legendre polynomials, we find
\beq\label{eq: cf-pk-leg-def}
    \xi(\vx,\vy) = \frac{1}{2}\sum_{\ell=0}^\infty\int_{\vk}e^{i\vk\cdot(\vx-\vy)}P_\ell(k)\left[L_\ell(\hk\cdot\hx)+(-1)^\ell L_\ell(\hk\cdot\hy)\right],
\eeq
where moments with $\ell>0$ are induced by redshift-space distortions and odd $\ell$ are sourced only by wide-angle effects (which vanish in the distant observer limit).\footnote{For connection to previous wide-angle effect literature \citep[e.g.,][]{2018MNRAS.476.4403C}, we note that, for $\ell=1$, $L_\ell(\hk\cdot\hx)+(-1)^\ell L_\ell(\hk\cdot\hy)\sim (s/x)\hk\cdot\hs$, where $\vs = \vx-\vy$ and $s\ll x,y$; as such, $P_{\ell=1}$ gives the first-order-in-$(s/x)$ correction to the power spectrum induced by our choice of the line-of-sight (but not the window function, since our estimators are window-deconvolved).} Finally, we can rewrite the power spectrum multipoles $P_\ell(k)$ in terms of band-power coefficients, $p_\alpha$ via $P_\ell(k) \approx \sum_b p_{\alpha}\Theta_b(k)$, where $\alpha\equiv\{b,\ell\}$ labels the coefficient and $\Theta_b(k)$ are a set of orthogonal top-hat functions encoding each $k$-bin. As such, the bandpower derivative appearing in \eqref{eq: gen-estimator-Sinv} is given by
\beq\label{eq: xi-p-alpha-deriv}
    \boxed{\frac{\partial{\xi(\vx,\vy)}}{\partial p_\alpha} = \frac{1}{2}\int_{\vk}e^{i\vk\cdot(\vx-\vy)}\Theta_b(k)\left[L_\ell(\hk\cdot\hx)+(-1)^\ell L_\ell(\hk\cdot\hy)\right]}.
\eeq
This is the key quantity required to implement the quasi-optimal estimators.

\subsection{General Estimators}
\noindent Following the discussion in \S\ref{subsec: gen-estimators}, the general quadratic estimator for the power spectrum coefficient $p_\alpha$ is given in matrix form as
\begin{empheq}[box=\widefbox]{align}\label{eq: pk-estimator-def}
    \hat{p}_\alpha &\equiv \sum_\beta \F_{\alpha\beta}^{-1}\left[p_\beta^{\rm num}-p_{\beta}^{\rm bias}\right]\\\nonumber
    p_\alpha^{\rm num} &= \frac{1}{2}[\Si d]^\dag\cdot\frac{\partial\xi}{\partial p_\alpha}\cdot[\Si d ]\\\nonumber
    p_\alpha^{\rm bias} &= \resub{\frac{1}{2}\mathrm{Tr}\left(\frac{\partial\xi}{\partial p_\alpha}\cdot[\Si \N\Sit]\right)}\\\nonumber
    \F_{\alpha\beta} &= \frac{1}{2}\mathrm{Tr}\left(\frac{\partial \xi}{\partial p_\alpha}\cdot[\Si \P]\cdot\frac{\partial\xi}{\partial p_\beta}\cdot[\Si \P]^\dag\right)
\end{empheq}
(cf.\,\ref{eq: gen-estimator-Sinv}, though with complex conjugates chosen to such that $\xi_{ij}=\av{\delta_i\delta_j^*}$ rather than $\av{\delta_i\delta_j}$), \resub{where the noise term is defined in \eqref{eq: two-point-noise}.} The quasi-optimal estimator \eqref{eq: pk-estimator-def} contains two key pieces: a numerator, $p_\alpha^{\rm num}$, which is quadratic in the data, encoding the standard power spectrum estimator (essentially computing a weighted version of $\int_{\vk}|d(\vk)|^2$), and a data-independent Fisher matrix, $\F_{\alpha\beta}$, which both normalizes the estimator and removes any leakage induced by the window function. 

As discussed in \S\ref{sec: estimator-theory}, the estimator is optimal and saturates its Cram\'{e}r-Rao bound if (a) the data is Gaussian distributed and (b) the $\Si$ weighting satisfies \eqref{eq: general-Sinv-optimality}. For general $\Si$, the Gaussian covariance is specified by
\beq
    \resub{\mathrm{cov}\left([\F\hat{p}]_\alpha,[\F\hat{p}]_{\beta}\right) = \frac{1}{2}\mathrm{Tr}\left(\frac{\partial\xi}{\partial p_\alpha}\cdot[\Si \C \Sit]\cdot\frac{\partial \xi}{\partial p_\beta}\cdot [\Si\C \Sit]\right)},
\eeq
for pixel-space covariance $\C_{ij} = \av{d_id^*_j}$ with $\mathrm{cov}\left(\hat{p}_\alpha,\hat{p}_{\beta}\right)=\F_{\alpha\beta}^{-1}$ in the optimal limit.\footnote{It is interesting to note that estimators with reduced variance are possible if the likelihood of $d$ is non-Gaussian (\textit{i.e.}\ if the fiducial $(n>2)$-point functions in \eqref{eq: edgeworth} are non-zero). In this case, one adds higher-order terms to the above estimators (without inducing bias) depending on the fiducial higher-point functions. In practice, these effects are small unless one is interested in very small scales (whence the non-Gaussianity in $\delta$ is large) or with very high shot-noise (when that in $\epsilon$ is significant). Such contributions will not be considered in this work though have been discussed in \citep{2021PhRvD.103j3504P}.} Finally, we note that the estimator is unbiased as long as our parametrization of $\xi(\vx-\vy)$ is complete, \textit{i.e.}\ if we capture all anisotropies (and beyond) of the two-point function. For example, if the underlying Universe contains only a monopole with $P(\vk;\hx)=P(k)$, the $\ell=0$ estimator is unbiased. However, one \textit{can} include higher-order terms in the estimator, e.g., $\ell=2$ modes; these can reduce the variance slightly (due to window-function induced leakage), but will not change the bias properties. This further implies that excluding odd-$\ell$ terms in the estimator does not induce bias at leading order (since these are suppressed by $|\vx-\vy|/x\ll 1$, and we always fully account for any contributions from the window functions).

Finally, we note one extension of the power spectrum estimators presented above. A crucial assumption of our formalism has been that the two-point function of the overdensity, $\xi$, can be represented by a reasonably small number of basis coefficients: the band-powers, $p_\alpha$, and that these can be efficiently predicted by some theoretical model. If the $k$-space bins are wide or the mask-induced leakage is severe, such assumptions may break down, at which point it is desirable to forward model the effects of the mask (\textit{i.e.}\ comparing window-convolved theory to windowed data), rather than inverting it (\textit{i.e.}\ comparing the raw theory to `unwindowed' data). By extending the above formalism, one can account for such effects (which essentially require an understanding of how to bin-integrate the theory) by computing also a `binning matrix', $\G_{\alpha\iota}$, which allows a finely-binned set of theory bandpowers, $p^{\rm fine}_\iota$ to be translated to the densely-binned observed quantities $\hat{p}_\alpha$. As we discuss in Appendix \ref{app: binning-theory}, the relevant matrix can be computed numerically, resulting in a procedure akin to the \textit{pseudo}-$C_\ell$ estimators \resub{used in the CMB community} \citep{Alonso:2018jzx} (see also the bin-convolution matrices of \citep{Beutler:2021eqq}). Whilst this picture is formally desirable, it is extremely expensive to compute analogous `bin convolution' matrices for statistics beyond the power spectrum, thus in this work, we principally consider window-deconvolved estimators, adopting a relatively fine (and extensive) binning to limit the above issues.

\subsection{Practical Implementation}\label{subsec: pk-implementation}
\noindent We now discuss how the estimators of \eqref{eq: pk-estimator-def} can be implemented in practice, remaining agnostic of the weighting scheme $\Si$ (with various choices discussed in \S\ref{subsec: weighting}). Details on the particular code implementation will be presented in \S\ref{sec: implementation}.

\subsubsection{Numerator}
\noindent From \eqref{eq: xi-p-alpha-deriv}, the numerator of the power spectrum estimator can be written
\beq
    p^{\rm num}_\alpha &=& \frac{1}{2}\int_{\vk}\Theta_b(k)\left(\int {\rm d}\vx\,e^{i\vk\cdot\vx}L_\ell(\hk\cdot\hx)[\Si d](\vx)\right)\left(\int {\rm d}\vy\,e^{-i\vk\cdot\vy}[\Si d](\vy)\right),
\eeq
symmetrizing over the two permutations. We note that $p^{\rm num}_\alpha$ is explicitly real (imaginary) for even (odd) $\ell$.\footnote{In practice, we take the imaginary part of the odd-$\ell$ estimator to keep all quantities real. This is equivalent to replacing $p_\alpha\to ip_\alpha$ in the original power spectrum definition if $\ell$ is odd.} In the distant observer limit $L_\ell(\hk\cdot\hx)\to L_\ell(\hk\cdot\hn)$, thus the numerator takes the simple form
\beq
    \left.p^{\rm num}_\alpha\right|_{\rm global}= \frac{1}{2}\int_{\vk}\Theta_b(k)L_\ell(\hk\cdot\hn)\left|\Si d\right|^2(\vk)
\eeq
for even $\ell$ and vanishes otherwise. This is the standard (unnormalized) quadratic estimator. Outside this regime, we expand the Legendre polynomial in normalized real spherical harmonics as $L_\ell(\hk\cdot\hx) = \sum_{m}\bar{Y}_{\ell m}(\hk)\bar{Y}_{\ell m}(\hx)$,\footnote{Explicitly $\sqrt{(2\ell+1)/4\pi}\bar{Y}_{\ell m} = (i/\sqrt{2})\left[Y_{\ell m}-(-1)^mY_{\ell(-m)}\right]$ if $m<0$, $(1/\sqrt{2})\left[Y_{\ell(-m)}+(-1)^mY_{\ell m}\right]$ if $m>0$, and $Y_{\ell 0}$ if $m=0$, where $Y_{\ell m}$ is the standard (complex) spherical harmonic.} giving
\beq
    p^{\rm num}_\alpha = \frac{1}{2}\sum_m\int_{\vk}\Theta_b(k)\bar{Y}_{\ell m}(\hk)\left(\int {\rm d}\vx\,e^{i\vk\cdot\vx}\bar{Y}_{\ell m}(\hx)[\Si d](\vx)\right)[\Si d](\vk).
\eeq
This is \resub{equivalent to} the usual Yamamoto estimator \citep[e.g.,][]{2017JCAP...07..002H,2006PASJ...58...93Y}, \resub{applied to a filtered field $\Si d$ (which includes any necessary deconvolution of the pixel-window)}. Denoting the (fast) Fourier transform by $\mathrm{FT}$, this is given explicitly by
\beq\label{eq: pk-num-fft}
    \textcolor{red}{\boxed{p^{\rm num}_\alpha = \frac{1}{2}\sum_m\int_{\vk}\Theta_b(k)\bar{Y}_{\ell m}(\hk)\left(\mathrm{FT}[\bar{Y}_{\ell m}\Si d](\vk)\right)^*\mathrm{FT}[\Si d](\vk),}}
\eeq
which requires $(2\ell+1)$ FFTs for all $k$-bins at a single $\ell$, and a Fourier-space sum for each $\alpha$ component. If computation of $\Si$ is rate-limiting (which may be true for optimal weights, implemented via a conjugate-gradient descent pipeline), the scaling is set by the number of $\Si$ applications; here, we require just one.

\subsubsection{Fisher Matrix \& Bias}\label{subsubsec: pk-fish-computation}
\noindent The Fisher matrix and bias terms in \eqref{eq: pk-estimator-def} are more difficult to compute, since they involve the trace over an $N_{\rm pix}\times N_{\rm pix}$-dimensional matrix. Whilst this is heinously expensive to compute explicitly, the trace can be estimated numerically by invoking the Girard-Hutchinson estimator \citep{girard89,hutchinson90} (as in \citep{2021PhRvD.103j3504P,2011MNRAS.417....2S,Philcox:2023uwe}). For a general high-dimensional matrix $\mathsf{M}$, this computes the trace as a Monte Carlo sum:
\beq\label{eq: girard-hutchinson}
    \mathrm{Tr}\left(\mathsf{M}\right) = \mathbb{E}_{\omega}[\omega^\dag\cdot \mathsf{M}\cdot\omega] \approx \frac{1}{N_{\rm mc}}\sum_{n=1}^{N_{\rm mc}}\left(\omega^{(n)}\right)^\dag\cdot \mathsf{M}\cdot\omega^{(n)},
\eeq
where $\omega^{(n)}$ are a set of $N_{\rm mc}$ \resub{independent and identically distributed (iid)} $N_{\rm pix}$-dimensional vectors satisfying $\mathbb{E}_\omega[\omega\omega^\dag]=\mathsf{I}$ (usually standardized Gaussian random fields or binary variables with $\omega\sim\mathrm{Unif}(\pm1)^{N_{\rm pix}}$). Importantly, this allows the trace to be computed without explicit knowledge of $\mathsf{M}$, only its action on an arbitrary map. Whilst generalizations to this formalism exist with faster convergence \citep[e.g.,][]{meyer2021hutch,Epperly_2024}, we will here adopt the simpler Girard-Hutchinson procedure due to its factorization properties.

Using \eqref{eq: girard-hutchinson}, we may rewrite the power spectrum bias term and Fisher matrix as an expectation over a set of Gaussian random maps $a$ with known covariance $\mathsf{A}\equiv\av{aa^\dag}$:
\beq
    p^{\rm bias}_\alpha &=& \resub{\frac{1}{2}\av{\left(\Si a\right)^\dag\cdot\frac{\partial\xi}{\partial p_\alpha}\cdot\left(\Si\N\Ai a\right)}_a}, \qquad
    \mathcal{F}_{\alpha\beta} = \frac{1}{2}\av{\left(\Si \P a\right)^\dag\cdot\frac{\partial\xi}{\partial p_\alpha}\cdot\left(\Si \P\right)\cdot\frac{\partial \xi}{\partial p_\beta}\cdot\Ai a}_a;
\eeq
as above, these can be estimated by a Monte Carlo summation over $N_{\rm mc}$ realizations, with an asymptotic error scaling of $\mathcal{O}(N_{\rm mc}^{-1})$, independent of the dimension: this is much faster than obtaining the matrix from the ideal covariance of $p^{\rm num}_\alpha$, and typically requires a few tens of iterations to \resub{ensure that the bias is significantly} below the noise threshold. Defining $\mathsf{Q}_\alpha[u] \equiv \partial\xi/\partial p_\alpha\cdot u$, the bias and Fisher matrix can be written \resub{in factorized form:
\beq
    p^{\rm bias}_\alpha &=& \frac{1}{2}\av{\left(\Si  \N\Ai a\right)^\dag\cdot\mathsf{Q}_\alpha[\Si a]}^*_a, \qquad \mathcal{F}_{\alpha\beta} = \frac{1}{2}\av{\left(\P^\dag\Sit\mathsf{Q}_\alpha[\Si \P a]\right)^\dag\cdot\mathsf{Q}_\beta[\Ai a]}_a,
\eeq
which avoids the na\"ive $\mathcal{O}(N_{\rm pix}^2$) scaling}. Notably, $\mathsf{Q}_\alpha[u]$ can be written in terms of Fourier transforms (from \ref{eq: xi-p-alpha-deriv}): 
\beq\label{eq: Q-alpha-local}
    \mathsf{Q}_\alpha[u](\vx) &=& \frac{1}{2}\sum_m\left\{\bar{Y}_{\ell m}(\hx)\int_{\vk}e^{i\vk\cdot\vx}\Theta_b(k)\bar{Y}_{\ell m}(\hk)u(\vk)+(-1)^\ell\int_{\vk}e^{i\vk\cdot\vx}\Theta_b(k)\bar{Y}_{\ell m}(\hk)\int {\rm d}\vy\,e^{-i\vk\cdot\vy}\bar{Y}_{\ell m}(\hy)u(\vy)\right\}\\\nonumber
    &=& \frac{1}{2}\sum_m\left\{\bar{Y}_{\ell m}(\hx)\mathrm{IFT}\left[\Theta_b\bar{Y}_{\ell m}\mathrm{FT}[u]\right](\vx)+(-1)^\ell\mathrm{IFT}\left[\Theta_b\bar{Y}_{\ell m}\mathrm{FT}[\bar{Y}_{\ell m}u]\right](\vx)\right\},
\eeq
and computed with $3(2\ell+1)$ FFTs for a given $k$-bin. In the distant-observer limit, we can further write
\beq\label{eq: Q-alpha-global}
    \left.\mathsf{Q}_\alpha[u](\vk)\right|_{\rm global} &=& \Theta_b(k)L_\ell(\hk\cdot\hn)u(\vk)
\eeq
for even $\ell$, vanishing else. This requires only one FFT.

\resub{The above expressions can be simplified further by noting that, due to symmetry, we can replace some of the $\mathsf{Q}$ factors by their asymmetric equivalents (defining $\Q = (\Q^{\rm asym}+\Q^{{\rm asym},\dag})/2$). This yields the final form:
\begin{empheq}[box=\widefboxred]{align}\label{eq: fish-from-Q}
    \textcolor{red}{p^{\rm bias}_\alpha} &\textcolor{red}{= \frac{1}{2}\av{\left(\Si  \N\Ai a\right)^\dag\cdot\mathsf{Q}^{\rm asym}_\alpha[\Si a]}^*_a}\\\nonumber
    \textcolor{red}{\mathcal{F}_{\alpha\beta}} &\textcolor{red}{= \frac{1}{2}\av{\left(\P^\dag\Sit\mathsf{Q}_\alpha[\Si \P a]\right)^\dag\cdot\mathsf{Q}^{\rm asym}_\beta[\Ai a]}_a,}
\end{empheq}
where
\beq
    \mathsf{Q}^{\rm asym}_\alpha[u](\vk) &=& \Theta_b(k)(-1)^\ell\sum_m\bar{Y}_{\ell m}(\hk)\int {\rm d}\vy\,e^{-i\vk\cdot\vy}\bar{Y}_{\ell m}(\hy)u(\vy)=\Theta_b(k)(-1)^\ell\sum_m\bar{Y}_{\ell m}\mathrm{FT}[\bar{Y}_{\ell m}u](\vk),
\eeq
in Fourier-space.} This is a particularly powerful result, since all $\Q^{\rm asym}_\alpha[u]$ maps for a given $u$ can be computed with just ($\ell_{\rm max}+1)^2$ FFTs (for all $\ell\leq \ell_{\rm max}$), regardless of the number of $k$-bins. 




In the above formalism, we compute the Fisher matrix as follows (with the shot-noise term computed similarly):
\begin{enumerate}
    \item Draw a random map $a$ from a known covariance matrix $\A$.
    \item Filter the map by $\Si\P$ and $\Ai$, both of which can be usually applied using FFTs. 
    \item \resub{Compute the weighted maps $(-1)^\ell\sum_m\bar{Y}_{\ell m}(\hk)\int {\rm d}\vy\,e^{-i\vk\cdot\vy}\bar{Y}_{\ell m}(\hy)[\Ai a](\vy)$ for each $\ell$ (unless working in the distant-observer limit).}
    \item For each of the $N_{\rm bins}$ choices of $\alpha$:
    \begin{enumerate}
        \item Compute $\mathsf{Q}_\alpha[\Si \P a]$, using FFTs, as in \eqref{eq: Q-alpha-local} or \eqref{eq: Q-alpha-global}.
        \item \resub{Weight by $\P^\dagger\Sit$, and transform to Fourier-space.}
        \item \resub{Compute the dot-product with $\Q_\beta^{\rm asym}[\Ai a]$ as a multiplication in Fourier-space (without FFTs).}
    \end{enumerate} 
    \item Iterate the above steps over $N_{\rm mc}$ Monte Carlo realizations to form the trace estimates. 
\end{enumerate}
\resub{Assuming that the FFTs are rate-limiting, the} Fisher matrix algorithm has complexity $\mathcal{O}(N_{\rm bins}N_{\rm mc}N_{\rm pix}\log N_{\rm pix})$, \resub{with $\mathcal{O}(N_{\rm mc}N_{\rm pix}\log N_{\rm pix})$ for the bias term}. Whilst this is slower than the estimator numerator (which involves only $(2\ell+1)$ FFTs in total), it is independent of the data and thus only has to be estimated once per choice of survey geometry and mask. \resub{Unlike the approach suggested in \citep{2021PhRvD.103j3504P,Philcox:2021ukg}, this approach does not require holding $\mathcal{O}(N_{\rm bins})$ maps in memory, which implies that it can easily be applied to large datasets.}

\subsubsection{Weighting Schemes}
\noindent Finally, we must specify a form for $\A$ and $\Si$. The above algorithm converges for any invertible $\A$, and, under certain assumptions, has an error independent of $\A$. In \polybin, we make the simple choice to generate periodic Gaussian random fields $a(\vx)$ with power spectrum $P_A(\vk) = [P_0(k)+1/\bar{n}]$ \resub{for some fiducial power spectrum monopole $P_0(k)$} (close to that of $d$); this has the trivial inverse
\beq
    \Ai(\vx,\vy) = \int_{\vk}e^{i\vk\cdot(\vx-\vy)}\frac{1}{P_A(\vk)}.
\eeq
This approach differs slightly from previous works which used lognormal $a$ maps \citep{2021PhRvD.103j3504P} or assumed $\Ai = \tCi$, \resub{for approximate inverse covariance $\tCi$} \citep{Philcox:2021ukg}. \resub{To compute the shot-noise, we instead use $P_A(\vk) = 1/[P_0(k)+1/\bar{n}]$, which is found to yield faster convergence.}

For $\Si$, we may either assume a simplified weighting (e.g., \ref{eq: Sinv-ideal} or \ref{eq: Sinv-fkp}), or numerically compute the full conjugate-gradient-descent solution, as discussed in \S\ref{subsec: weighting}. To implement the latter, we must rewrite \eqref{eq: cgd-operation-eqn} in terms of the fiducial power spectrum multipoles $P_\ell(k)$; with our definitions of $\xi$ \eqref{eq: cf-pk-leg-def}, we find
\beq\label{eq: cgd-operation-eqn}
    \frac{1}{2}n(\vx)\int {\rm d}\vy\int_{\vk}e^{i\vk\cdot(\vx-\vy)}\sum_{\ell=0}^{\ell_{\rm max}} P_\ell(k)\left[L_\ell(\hk\cdot\hx)+(-1)^\ell L_\ell(\hk\cdot\hy)\right][\Phi\Si v](\vy)+\frac{n_2(\vx)}{n(\vx)}[\Phi\Si v](\vx)= [\mathsf{M}^{-1} v](\vx),
\eeq
where $\Phi(\vx)=0$ if $n(\vx)=0$ and unity else, and $v$ is the map that we wish to apply $\Si$ to. Expanding the Legendre polynomials as before, we arrive at the form
\beq\label{eq: Sinv-optimal}
    &&\frac{1}{2}\sum_{\ell m}n(\vx)\bar{Y}_{\ell m}(\hx)\left\{\mathrm{IFT}\left[P_\ell\bar{Y}_{\ell m}\mathrm{FT}[\Phi\Si v]\right](\vx)+(-1)^\ell \mathrm{IFT}\left[P_\ell \bar{Y}_{\ell m}\mathrm{FT}[\bar{Y}_{\ell m}\Phi \Si v]\right](\vx)\right\} + \frac{n_2(\vx)}{n(\vx)}[\Phi\Si v](\vx)\\\nonumber
    &&= [\mathsf{M}^{-1}v](\vx),
\eeq
which is a sequence of $3(\ell_{\rm max}+1)^2-1$ chained FFTs. In the distant observer limit, this simplifies significantly:
\beq 
    n(\vx)\mathrm{IFT}\left[\sum_\ell P_\ell L_\ell\mathrm{FT}[\Phi\Si v]]\right](\vx)+\frac{n_2(\vx)}{n(\vx)}\Phi(\vx)[\Si v](\vx) = [\mathsf{M}^{-1}v](\vx),
\eeq
\resub{for even $\ell$, requiring only $2$ FFTs.} \resub{We can similarly obtain an equation for $\Sit$, with the final result
\beq
    &&\frac{1}{2}\sum_{\ell m}\Phi(\vx)\left\{\bar{Y}_{\ell m}(\hx)\mathrm{IFT}\left[P_\ell\bar{Y}_{\ell m}\mathrm{FT}[n\Sit v]\right](\vx)+(-1)^\ell\mathrm{IFT}\left[P_\ell\bar{Y}_{\ell m}\mathrm{FT}[\bar{Y}_{\ell m}n\Sit v]\right](\vx)\right\}+\frac{n_2(\vx)}{n(\vx)}[\Phi\Sit v](\vx)\\\nonumber
    &&= [\mathsf{M}^{-1}v](\vx),
\eeq
or 
\beq
    &&\frac{1}{2}\Phi(\vx)\mathrm{IFT}\left[\sum_\ell P_\ell L_\ell\mathrm{FT}[n\Sit v]\right](\vx)+\frac{n_2(\vx)}{n(\vx)}[\Phi\Sit v](\vx)= v(\vx)
\eeq
in the distant-observer limit.} Using the conjugate-gradient-descent algorithm coupled with an appropriate \resub{(Hermitian)} approximation for $\Si,\Sit$ (acting as a preconditioner) these equations can be efficiently solved to compute $[\Si v]$ \resub{and $[\Sit v]$ for arbitrary $v$. We typically adopt a smooth approximation to $n(\vx),n_2(\vx)$ when computing the optimal weights (but can use the masks computed from random fields to define the Fisher matrix, \textit{i.e.}\ the mask deconvolution).}

The conclusion of the above discussion is that the formal power spectrum estimators derived in \S\ref{sec: estimator-theory} can be practically implemented on observational or simulated data. This is made possible by two key factors: (1) that the bandpower derivative $\partial \xi/\partial p_\alpha$ can be efficiently applied to maps using FFTs; (2) we can rewrite the trace terms as Monte Carlo summations. Under certain assumptions (namely that the two-point function is fully described by the power spectrum bandpowers), we can thus obtain an efficient and unbiased estimator for $p_\alpha$. In the next section, we will discuss how the above forms are simplified the ideal scenario (relevant to periodic-box simulations) and give a comparison to the standard (window-convolved) forms.

\subsection{Ideal Limits}
\subsubsection{Uniform Density}
\noindent When computing the power spectrum bandpowers from $N$-body simulations, one is usually interested in the ideal limit of a uniform and everywhere-defined background density $n(\vx) = n_2(\vx) = \bar{n}$ with redshift space distortions implemented along a global line-of-sight, $\hn$ (which allows restriction to even $\ell$). By translation invariance, the power spectrum weighting scheme ought to be diagonal in harmonic-space, thus we can set $\Si(\vx,\vy) = \int_{\vk}e^{i\vk\cdot(\vx-\vy)}1/S(\vk)$; from \eqref{eq: Sinv-ideal}, the optimal solution is \resub{$S(\vk) = m(\vk)\bar{n}\left[P(\vk)+1/\bar{n}\right]\equiv m(\vk)\bar{n}P_{\rm fid}(\vk)$} for \resub{and pixelation window $m(\vk)$} and fiducial power spectrum $P_{\rm fid}(\vk)$, which includes Poissonian shot-noise. 

Under the above assumptions, the power spectrum estimators of \eqref{eq: pk-estimator-def} simplify to
\beq\label{eq: pk-limit}
    &&\textcolor{blue}{\boxed{\left.p^{\rm num}_\alpha\right|_{\rm ideal} = \frac{1}{2}\frac{1}{\bar{n}^2}\int_{\vk}\Theta_b(k)L_\ell(\hk\cdot\hn)\frac{1}{P^2_{\rm fid}(\vk)}\left|\frac{d(\vk)}{m(\vk)}\right|^2}}\\\nonumber
    &&\left.p^{\rm bias}_\alpha\right|_{\rm ideal} = \frac{1}{2}\frac{1}{\bar{n}}\int_{\vk}\Theta_b(k)L_\ell(\hk\cdot\hn)\frac{1}{P^2_{\rm fid}(\vk)}\\\nonumber
    &&\left.\mathcal{F}_{\alpha\beta}\right|_{\rm ideal} = \frac{1}{2}\int_{\vk}\Theta_b(k)\Theta_{b'}(k)L_\ell(\hk\cdot\hn)L_{\ell'}(\hk\cdot\hn)\frac{1}{P^2_{\rm fid}(\vk)}.
\eeq
The numerator is the standard quadratic power spectrum estimator applied to \resub{pixel-window-deconvolved and inverse-variance}-filtered data. Assuming isotropic $P_{\rm fid}$ and working in the continuous limit (ignoring any discreteness artefacts from a finite $\vk$-space grid), $\F_{\alpha\beta}$ is diagonal in both $b$ and $\ell$:
\beq\label{eq: fish-ideal-continuous}
    \textcolor{blue}{\boxed{\left.\F_{\alpha\beta}\right|_{\rm ideal} \to \frac{1}{2}\frac{\delta_{\alpha\beta}^{\rm K}}{2\ell+1}\int_{\vk}\frac{\Theta_b(k)}{P^2_{\rm fid}(k)},}}
\eeq
which is simply the bin volume, weighted by $P_{\rm fid}^{-2}(k)$. In general, this also captures couplings between different $\ell$ modes induced by anisotropic weightings and the finite size of Fourier-space pixels (whence $\int {\rm d}\hk\,L_\ell(\hk\cdot\hn)L_{\ell'}(\hk\cdot\hn)\not\propto\delta^{\rm K}_{\ell\ell'}$). Finally, the shot-noise term $p^{\rm bias}$ contains only contributions from $\ell=0$ (with a net contribution to $\hat{p}_\alpha$ of $1/\bar{n}$, as expected) again assuming the continuous limit and isotropic $P_{\rm fid}$.

\subsubsection{FKP Weights}
\noindent An additional limit of interest is obtained by fixing $\Si$ to the FKP form given in \eqref{eq: Sinv-fkp} \citep[cf.][]{1994ApJ...426...23F}. \resub{As discussed in \S\ref{subsec: weighting}, the relevant weighting depends on how the datavector, $d$, is constructed: if we apply the FKP weights at the catalog level, we weight according to $\mathsf{M}^{-1}d$, whilst if we do not, we weight the data by $[1+n(\vx)P_{\rm FKP}]^{-1}\mathsf{M}^{-1} = w_{\rm FKP}(\vx)\mathsf{M}^{-1}$. Assuming the former option, as well as} the distant-observer limit for simplicity, this leads to the following estimators:
\beq\label{eq: pk-fkp-lim}
    \left.p^{\rm num}_\alpha\right|_{\rm FKP} &=& \frac{1}{2}\int_{\vk}\Theta_b(k)L_\ell(\hk\cdot\hn)\left|\int {\rm d}\vx\,e^{-i\vk\cdot\vx}[\mathsf{M}^{-1}d](\vx)\right|^2\\\nonumber
    \left.p^{\rm bias}_\alpha\right|_{\rm FKP} &=& \frac{1}{2}\int_{\vk}\Theta_b(k)L_\ell(\hk\cdot\hn)\int {\rm d}\vx\,n_2(\vx)\\\nonumber
    \left.\mathcal{F}_{\alpha\beta}\right|_{\rm FKP} &=& \frac{1}{2}\int_{\vk\,\vk'}\Theta_b(k)\Theta_{b'}(k')L_\ell(\hk\cdot\hn)L_{\ell'}(\hk'\cdot\hn)\left|\int {\rm d}\vx\,n(\vx)e^{i(\vk'-\vk)\cdot\vx}\right|^2,
\eeq
\resub{where $n_2$ is the doubly-weighted mask defined in \eqref{eq: two-point-noise}, which includes two copies of the FKP weight.} The numerator and shot-noise matches that used in standard FKP power spectrum estimators (for both line-of-sight choices) \citep[e.g.,][]{2017JCAP...07..002H,2015PhRvD..92h3532S,2015MNRAS.453L..11B,2006PASJ...58...93Y,1994ApJ...426...23F}, including that of the \textsc{nbodykit} code \citep{2018AJ....156..160H} (which uses an ungridded form of $p^{\rm bias}$, dropping contributions with $\ell>0$). Our normalization, $\F_{\alpha\beta}$, differs from that of the standard FKP estimator, which uses
\beq
    \mathcal{F}^{\rm conv}_{\alpha\beta} &=& \frac{1}{2}\int_{\vk}\Theta_b(k)\Theta_{b'}(k)L_\ell(\hk\cdot\hn)L_{\ell'}(\hk\cdot\hn)\int {\rm d}\vx\,n^2(\vx);
\eeq
up to discreteness effects, this leads to the full estimator
\beq
    \left.\hat{p}^{\rm conv}_\alpha\right|_{\rm FKP} &=& \frac{2\ell+1}{\int_{\vk}\Theta_b^2(k)}\left[\int {\rm d}\vx\,n^2(\vx)\right]^{-1}\int_{\vk}\Theta_b(k)L_\ell(\hk\cdot\hn)\left|\int {\rm d}\vx\,e^{-i\vk\cdot\vx}[\mathsf{M}^{-1}d](\vx)\right|^2.
\eeq
The difference between the two forms arises since the FKP method estimates `windowed' power spectra, which must be compared to mask-convolved theory spectra. In our estimators, the effects of bin-convolution are captured by the Fisher matrix, which involves the power spectrum of the mask itself; $\left|n(\vk-\vk')\right|^2$. Assuming that our parametrization of $\xi$ is complete, the two approaches are formally equivalent, \resub{differing only by a rotation}.

\section{Bispectrum Estimators}\label{sec: bispectra}
\subsection{Definitions}
\noindent Having derived quasi-optimal and (relatively) easy-to-implement power spectrum estimators, we now turn to the bispectrum. A central ingredient of the estimators discussed in \S\ref{subsec: spec-estimators} is the pixel-space three-point function $\av{d^id^jd^k}_c$: using the data model of \eqref{eq: data-model}, this is given by
\beq
    \av{d^{i_1}d^{i_2}d^{i_3}}_c = \P^{i_1j_1}\P^{i_2j_2}\P^{i_3j_3}\zeta_{j_1j_2j_3}+\N^{i_1i_2i_3},
\eeq
where $\zeta^{i_1i_2i_3}\equiv\av{\delta^{i_1}\delta^{i_2}\delta^{i_3}}_c$ is the three-point function of the overdensity field, and $\N^{i_1i_2i_3}$ is a Poisson noise term \resub{(which includes contributions proportional to both $1/\bar{n}^2$ and $P(\vk)/\bar{n}$, as discussed below)}. 
 
To proceed, we must relate $\zeta^{i_1i_2i_3}$ to the quantity of interest: the redshift-space bispectrum multipoles. This is achieved by first writing the map-space three-point function in Fourier-space: 
\beq\label{eq: Bk-from-zeta}
    \zeta(\vx_1,\vx_2,\vx_3) = \int_{\vk_{123}=\vec 0}e^{i(\vk_1\cdot\vx_1+\vk_2\cdot\vx_2+\vk_3\cdot\vx_3)}B(\vk_1,\vk_2,\vk_3;\hx_1,\hx_2,\hx_3),
\eeq
where we allow for explicit dependence on the three lines-of-sight $\hx_{1,2,3}$ and notate $\int_{\vk_{123}=\vec 0} = \int_{\vk_1\,\vk_2\,\vk_3}\delD{\vk_1+\vk_2+\vk_3}$, with the Dirac delta arising from translation invariance. As in \citep{Ivanov:2023qzb} (which follows \citep{2015PhRvD..92h3532S,Yankelevich:2018uaz,Rizzo:2022lmh}), we parametrize the anisotropy by a Legendre polynomial expansion about the longest $\vk$-vector:
\beq\label{eq: Bk-rsd}
    B(\vk_1,\vk_2,\vk_3;\hx_1,\hx_2,\hx_3) \approx \sum_{\ell=0}^\infty B_\ell(k_1,k_2,k_3)L_\ell(\hk_3\cdot\hx_3)\qquad (k_1\leq k_2\leq k_3),
\eeq
analogous to the Yamamoto form used in the power spectrum estimators (see \citep{Garcia:2020per} for more nuanced line-of-sight choices). Notably, this is not a complete expansion even in the distant observer limit (whence $\hx_1=\hx_2=\hx_3=\hn$); one should properly allow for the bispectrum to depend on two angles, which jointly parameterize orientation of the $\vk_1-\vk_2-\vk_3$ plane to the line-of-sight. Whilst this is formally possible (by expanding in spherical harmonics), it does not yield a separable expansion, as noted in \citep{2015PhRvD..92h3532S}, thus we here adopt the simplified form given in \eqref{eq: Bk-rsd}, which \citep{Gagrani:2016rfy} find to be roughly optimal. Alternative expansions also exist: \citep{2019MNRAS.484..364S,2017PhRvD..95f3508S,Byun:2022rvn} utilize a `BiPoSH' expansion allowing for multiple lines-of-sight (at the expense of many more components), whilst \citep{2020JCAP...06..041G} advocate for a double Legendre expansion (which is again non-separable).

Next, we must write the above expression in terms of the binned bispectrum coefficients, $\{b_\alpha\}$. As in \citep{Ivanov:2023qzb}, we define
\beq\label{eq: Bk-Yam}
    \boxed{B(\vk_1,\vk_2,\vk_3;\hx_1,\hx_2,\hx_3) \approx \sum_\alpha \frac{b_\alpha}{\Delta_\alpha}\left[\Theta_{b_1}(k_1)\Theta_{b_2}(k_2)\Theta_{b_3}(k_3)L_\ell(\hk_3\cdot\hx_3)+\text{5 perms.}\right],}
\eeq
where $\alpha\equiv\{b_1,b_2,b_3,\ell\}$ with $b_1\leq b_2\leq b_3$ indexes the bin, and we sum over the six triangle permutations of $\{(\vk_1,\hx_1),(\vk_2,\hx_2),(\vk_3,\hx_3)\}$, which ensure that the total bispectrum is symmetric under $(\vk_i,\hx_i)$ interchange.\footnote{For two-point functions, the square bracket in \eqref{eq: Bk-Yam} becomes $\Theta_{b_1}(k_1)\Theta_{b_2}(k_2)L_\ell(\hk_1\cdot\hx_1)+\Theta_{b_1}(k_2)\Theta_{b_2}(k_1)L_\ell(\hk_2\cdot\hx_2)$, recovering the previous definition since $\vk_1+\vk_2=\vec 0$.} Ignoring wide-angle corrections, we can restrict the above summation to even $\ell$. Finally, we must specify the `degeneracy factor', $\Delta_\alpha$, which is added to avoid double-counting, since our $k$-bins have finite width. Again following \citep{Ivanov:2023qzb}, this is given by
\beq\label{eq: Delta-alpha-def}
    \Delta_\alpha = \begin{cases} 1 & b_1\neq b_2\neq b_3\\ 2 & b_1=b_2 \neq b_3\\ 1+N_\ell(\vec b)/N_0(\vec b) & b_1\neq b_2=b_3\\ 2[1+2N_\ell(\vec b)/N_0(\vec b)] & b_1=b_2=b_3 \end{cases},
\eeq
for 
\beq\label{eq: N-ell-def}
    N_\ell(\vec b) = \int_{\vk_{123}=\vec 0}\frac{\Theta_{b_1}(k_1)\Theta_{b_2}(k_2)\Theta_{b_3}(k_3)}{P_{\rm fid}(k_1)P_{\rm fid}(k_2)P_{\rm fid}(k_3)}L_\ell(\hk_2\cdot\hk_3),
\eeq
where $P_{\rm fid}(k)$ is some fiducial power spectrum monopole, \resub{which we could optionally set to unity}. This has a complex form for $\ell>0$ due to the degeneracy in our bispectrum definition, \textit{i.e.}\ the fact that we parametrize anisotropy using only one of the three $\vk$-vectors. Formally, it can be obtained by asserting that the ideal bispectrum estimator (derived below, see also \citep{2015PhRvD..92h3532S})
\beq\label{eq: ideal-bk}
    \hat b^{\rm idealized}_\alpha \equiv \frac{2\ell+1}{N_0(\vec b)}\int_{\vk_{123}=\vec 0}\frac{\Theta_{b_1}(k_1)\Theta_{b_2}(k_2)\Theta_{b_3}(k_3)}{P_{\rm fid}(k_1)P_{\rm fid}(k_2)P_{\rm fid}(k_3)}L_\ell(\hk_3\cdot\hn)\,\times\,\delta(\vk_1)\delta(\vk_2)\delta(\vk_3)
\eeq
has expectation $b_\alpha$ in the limit of a global line-of-sight and no discreteness effects. In other words, we \textit{define} $b_\alpha$ by \eqref{eq: ideal-bk}; this can then be related to the full bispectrum by \eqref{eq: Bk-Yam}.

Collecting results, we find the following form for the three-point derivative used in \eqref{eq: gen-estimator-Sinv}:
\beq\label{eq: zeta-deriv}
    \boxed{\frac{\partial \zeta(\vx_1,\vx_2,\vx_3)}{\partial b_\alpha} = \frac{1}{\Delta_\alpha}\int_{\vk_{123}=\vec 0}e^{i(\vk_1\cdot\vx_1+\vk_2\cdot\vx_2+\vk_3\cdot\vx_3)}\left[\Theta_{b_1}(k_1)\Theta_{b_2}(k_2)\Theta_{b_3}(k_3)L_\ell(\hk_3\cdot\hx_3)+\text{5 perms.}\right].}
\eeq

\subsection{General Estimators}
\noindent The general bispectrum estimator follows from the discussion in \S\ref{subsec: gen-estimators}:
\begin{empheq}[box=\widefbox]{align}\label{eq: bk-estimator-def}
    \hat{b}_\alpha &\equiv \sum_\beta\F_{\alpha\beta}^{-1}[b_\beta^{\rm num}-b_{\beta}^{\rm bias}]\\\nonumber
    b_\alpha^{\rm num} &= \frac{1}{6}\frac{\partial\zeta^{i_1i_2i_3}}{\partial b_\alpha}\left([\Si d]_{i_1}[\Si d]_{i_2}[\Si d]_{i_3}-3[\Si d]_{i_1}[\resub{\Si \C\Sit}]_{i_2i_3}\right)^*\\\nonumber
    b_\alpha^{\rm bias} &= \resub{\frac{1}{6}\sum_j\frac{\partial\zeta^{i_1i_2i_3}}{\partial b_\alpha}\left[\Si_{i_1j_1}\Si_{i_2j_2}\Si_{i_3j_3}\N^{j_1j_2j_3}\right]^*}\\\nonumber
    \F_{\alpha\beta} &= \frac{1}{6}\frac{\partial\zeta^{i_1i_2i_3}}{\partial b_\alpha}[\Si \P]^*_{i_1j_1}[\Si \P]^*_{i_2j_2}[\Si \P]^*_{i_3j_3}\frac{\partial\zeta^{j_1j_2j_3*}}{\partial b_\beta},
\end{empheq}
assuming $d_i$ to be real in $b_\alpha^{\rm num}$. 

As for the power spectrum estimator, \eqref{eq: bk-estimator-def}, comprises two main pieces: a data-dependent numerator, $b_\alpha^{\rm num}$, and a data-independent Fisher matrix, $\F_{\alpha\beta}$, which provides the normalization (and removes leakage between bins). Notably, the numerator contains both a cubic and a linear term: the latter does not change the estimator mean (since $\av{\Si d} = 0$), but can remove bias on large-scales.\footnote{Our linear term differs from that of \citep{Philcox:2021ukg}; the former work used $-\frac{1}{2}(\partial\zeta^{i_1i_2i_3}/\partial b_\alpha) [\Si d]_{i_1}[\Si \P]_{i_2i_3}$, which is correct only if $\Si$ is optimal. Our form will lead to slightly improved large-scale variance-suppression for general $\Si$.} For a general weighting scheme $\Si$, the covariance of $b^{\rm num}_{\alpha}$ is given by
\beq
    \mathrm{cov}\left(b^{\rm num}_\alpha, b^{\rm num}_\beta\right) &=& \frac{1}{6}\left[\frac{\partial\zeta^{i_1i_2i_3*}}{\partial b_\alpha}[\Si \C \Sit]_{i_1j_1}[\Si \C \Sit]_{i_2j_2}[\Si \C\Sit]_{i_3j_3}\frac{\partial\zeta^{j_1j_2j_3}}{\partial b_\beta}\right]^*,
\eeq
which is equal to $\F_{\alpha\beta}$ if the weighting \resub{satisfies $\Si\P = \Si\C\Sit$} and the data is Gaussian distributed. In this regime, the estimator is optimal. Finally, the estimator is unbiased up to the terms dropped in the redshift-space expansion of \eqref{eq: Bk-rsd}, though we note that such terms cannot contribute in the ideal limit, since the underlying spherical harmonic basis functions are orthogonal. 

\subsection{Practical Implementation}
\noindent Na\"ive implementation of \eqref{eq: bk-estimator-def} is difficult, since it involves summation over trilinear operators with dimension $N_{\rm pix}\times N_{\rm pix}\times N_{\rm pix}$. However, this can be recast in manifestly separable form by invoking translation invariance and Monte Carlo summation (\citep[cf.][]{Philcox:2021ukg,Ivanov:2023qzb} building on the CMB algorithms of \citep{2011MNRAS.417....2S}). Below, we give details on this procedure in the general case of non-uniform geometries -- idealized limits (relevant to simulation computations) will be given in \S\ref{subsec: bk-ideal}.

\subsubsection{Preliminaries}
\noindent To implement the bispectrum estimator, we require an efficient method of computing the $\partial_\alpha\zeta$ derivative, when applied to a triplet of fields, \textit{i.e.} $\beta_\alpha[u,v,w]\equiv \partial_\alpha \zeta^{i_1i_2i_3}u^*_{i_1}v^*_{i_2}w^*_{i_3}$. From \eqref{eq: zeta-deriv}, this is given by
\beq\label{eq: beta-def1}
    \beta_\alpha[u,v,w] &=& \frac{1}{\Delta_\alpha}\int_{\vk_{123}=\vec 0}\Theta_{b_1}(k_1)\Theta_{b_2}(k_2)\Theta_{b_3}(k_3)u^*(\vk_1)v^*(\vk_2)\left(\int {\rm d}\vx_3\,e^{-i\vk_3\cdot\vx_3}L_\ell(\hk_3\cdot\hx_3)w(\vx_3)\right)^*+\text{5 perms.}
\eeq
Using $\delD{\vk_1+\vk_2+\vk_3}=\int {\rm d}\vr\,e^{i(\vk_1+\vk_2+\vk_3)\cdot\vr}$, this can be written in separable form:
\beq\label{eq: beta-def}
    \boxed{\beta_\alpha[u,v,w] =\frac{1}{\Delta_\alpha}\int {\rm d}\vr\,g_{b_1,0}[u](\vr)g_{b_2,0}[v](\vr)g_{b_3,\ell}[w](\vr)+\text{5 perms.},}
\eeq
defining
\beq\label{eq: g-ell-def}
    g_{b,\ell}[u](\vr) = \int_{\vk}e^{i\vk\cdot\vr}\Theta_b(k)\left(\int {\rm d}\vx\,e^{-i\vk\cdot\vx}L_\ell(\hk\cdot\hx)u(\vx)\right)^*.
\eeq
To compute the real-space maps $g_{b,\ell}$ we can expand the Legendre polynomials in (normalized real) spherical harmonics, as before:
\beq
    &&g_{b,\ell}[u](\vr) = \int_{\vk}e^{i\vk\cdot\vr}\Theta_b(k)\sum_m\bar{Y}_{\ell m}(\hk)\left(\int {\rm d}\vx\,e^{-i\vk\cdot\vx}\bar{Y}_{\ell m}(\hx)u(\vx)\right)^*\\\nonumber
    &&\left.g_{b,\ell}[u](\vr)\right|_{\rm global} = \int_{\vk}e^{i\vk\cdot\vr}\Theta_b(k)L_\ell(\hk\cdot\hn)\left(\int {\rm d}\vx\,e^{-i\vk\cdot\vx}u(\vx)\right)^*\\\nonumber
\eeq
with the distant observer limit given in the second line. This can be easily computed with $(2\ell+2)$ FFTs: 
\beq
    g_{b,\ell}[u](\vr) = \mathrm{IFT}\left[\Theta_b\sum_{m}\bar{Y}_{\ell m}\mathrm{FT}[\bar{Y}_{\ell m}u]^*\right](\vr),\qquad \left.g_{b,\ell}[u](\vr)\right|_{\rm global} = \mathrm{IFT}\left[\Theta_bL_\ell\,\mathrm{FT}[u]^*\right](\vr)
\eeq
or just $2$ for a global line-of-sight. The full $\beta_\alpha$ term can thus be computed by a summation in real-space. These tricks can also be used to compute the $N_\ell(\vec b)$ factors in \eqref{eq: N-ell-def}:
\beq
    N_\ell(\vec b) = \sum_m\int {\rm d}\vr\,\left(\int_{\vk_1}e^{i\vk_1\cdot\vr}\frac{\Theta_{b_1}(k_1)}{P_{\rm fid}(k_1)}\right)\left(\int_{\vk_2}e^{i\vk_2\cdot\vr}\frac{\Theta_{b_2}(k_2)}{P_{\rm fid}(k_2)}\bar{Y}_{\ell m}(\hk_2)\right)\left(\int_{\vk_3}e^{i\vk_3\cdot\vr}\frac{\Theta_{b_3}(k_3)}{P_{\rm fid}(k_3)}\bar{Y}_{\ell m}(\hk_3)\right).
\eeq
This requires $(\ell_{\rm max}+1)^2N_k$ FFTs to compute in total.

\subsubsection{Numerator}
\noindent The bispectrum numerator $b^{\rm num}_\alpha$ can be written in terms of the $\beta$ operator of \eqref{eq: beta-def}. Explicitly
\beq\label{eq: b-numerator-practical}
    \textcolor{red}{\boxed{b^{\rm num}_\alpha = \frac{1}{6}\bigg\{\beta_\alpha[\Si d,\Si d,\Si d]-\left(\av{\beta_\alpha[\Si d,\Si \Delta,\Si \Delta]}_\Delta+\text{2 perms.}\right)\bigg\}.}}
\eeq
Here, we have rewritten the linear term as an average over random fields $\Delta$ with covariance $\av{\Delta\Delta^\dag} = \C_\Delta\approx\C$. This can be efficiently computed as a Monte Carlo summation, with $\{\Delta\}$ being a suite of simulations or Gaussian random fields.\footnote{Note that our estimator is not biased if $\C_\Delta\neq \C$, but its variance generically increases.} If one does not have access to accurate simulations, the linear term could also be computed by more nuanced schemes based on the Girard-Hutchinson estimator. For example, one could replace
\beq
    \av{\beta_\alpha[\Si d,\Si \Delta,\Si\Delta]}_\Delta \quad \to \quad \av{\beta_\alpha[\Si d,\Si \C\Ai a,\Si M^{-1}a]}_a
\eeq
for random fields $a$ specified by invertible covariance $\A \equiv\av{aa^\dag}$; this requires a computable form for $\C$ (\S\ref{subsec: pk-implementation}), but not simulations drawn from it. Given that the linear term is usually small (and vanishes in the ideal limit), this effect has little impact in practice, thus we adopt the simpler procedure in this work.

Given $N_{\rm sim}=\mathcal{O}(10^2)$ simulated realizations of $\Delta$, computation of the numerator \eqref{eq: b-numerator-practical} for all $\alpha$ bins requires $\left[(\ell_{\rm max}+1)^2+N_k(\ell_{\rm max}/2+1)\right](N_{\rm sim}+1)$ FFTs (each of which has $\mathcal{O}(N_{\rm pix}\log N_{\rm pix}$ complexity), where $N_k$ is the number of $k$-bins per dimension (\textit{not} the total number of bispectrum bins); this drops to $N_k(\ell_{\rm max}/2+1)(N_{\rm sim}+1)$ in the distant-observer limit. Furthermore, it requires $N_{\rm sim}+1$ invocations of the (possibly expensive) filtering $\Si$. Clearly, computation can be significantly expedited if one drops the linear term; as shown in \citep{Philcox:2021ukg}, this is often an excellent approximation since the associated variance reduction is usually important only on ultra-large scales.

\subsubsection{Poisson Noise}\label{subsubsec: bk-shot}
\noindent To estimate the shot-noise term, we use a similar procedure as for the power spectrum. \resub{We first consider the explicit form for the noise term given a spectroscopic survey defined by a collection of galaxies and random points, as in \S\ref{subsec: spec-estimators}. Extracting all terms arising from Poisson noise and averaging over the underlying random fields, we find an expression similar to \eqref{eq: two-point-noise}:
\beq\label{eq: three-point-noise}
    \N(\vx,\vy,\vz) &=& \left(1+\alpha^3/\alpha_3\right)\int {\rm d\vw}\,m(\vx-\vw)m(\vy-\vw)m(\vz-\vw)\overline{n}_3(\vw)\\\nonumber
    &&\,+\,\bigg\{\int {\rm d}\vw\,m(\vy-\vw)m(\vz-\vw)\int {\rm d}\vw'\,m(\vx-\vw')\overline{n}_2(\vw)\overline{n}_1(\vw')\av{\delta_2(\vw)\delta_1(\vw')}+\text{2 perms.}\bigg\}
\eeq
defining the $k$-weighted overdensities and background densities $\delta_k$ and $\overline{n}_k$ respectively, with $\alpha_k = \sum_{i=1}^{N_g}w_{g,i}^k/\sum_{i=1}^{N_r}w_{r,i}^k$.\footnote{This can be defined by considering taking the expectation of $d(\vx)d(\vy)d(\vz)$ and isolating terms containing coincident galaxies or randoms.} This can be estimated using the data and random catalogs via
\beq
    \widehat{\N}(\vx,\vy,\vz) &=& -2\int {\rm d\vw}\,m(\vx-\vw)m(\vy-\vw)m(\vz-\vw)n_3(\vw)\\\nonumber
    &&\,+\,\bigg\{\int {\rm d}\vw\,m(\vy-\vw)m(\vz-\vw)\int {\rm d}\vw'\,m(\vx-\vw')d_2(\vw)d_1(\vw')+\text{2 perms.}\bigg\}
\eeq
defining the (unpixelized) weighted randoms and data:
\beq\label{eq: n3-def}
    n_3(\vx) &=& \left(\alpha_3+\frac{1}{2}\alpha^3+\frac{3}{2}\alpha_2\alpha\right)\sum_{i=1}^{N_r}w_{r,i}^3\delta_{\rm D}(\vx-\vx_{r,i})\\\nonumber
    d_2(\vx) &=& \sum_{i=1}^{N_g} w_{g,i}^2\delta_{\rm D}(\vx-\vx_{g,i}) -\alpha_2\sum_{i=1}^{N_r} w_{r,i}^2\delta_{\rm D}(\vx-\vx_{r,i}),
\eeq
where we explicitly correct for shot-noise contributions to $d_2(\vw)d_1(\vw')$. In the absence of a pixel window, and assuming a uniformly weighted sample, the shot-noise takes the simple limit
\beq
    \N(\vx,\vy,\vz) &\to& \overline{n}(\vx)\delta_{\rm D}(\vx-\vy)\delta_{\rm D}(\vx-\vz) + \left(\xi(\vx,\vy)\overline{n}(\vx)\overline{n}(\vy)\delta_{\rm D}(\vy-\vz)+\text{2 perms.}\right),
\eeq
which encompasses the standard $\bar{n}^{-2}$ and $\bar{n}^{-1}P(\vk)$ terms \citep[e.g.,][]{Ivanov:2021kcd}.}

\resub{To estimate these terms, we start from} \eqref{eq: bk-estimator-def}, introducing two pairs of iid random fields $\{a^{(1,2)}\}$ with covariance $\A$, \resub{as for the normalization matrix. Decomposing the noise into two terms, $\N_{ijk} = -2\N^{(3)}_{ijk}+(d_i\N^{(2)}_{jk}+\text{2 perms.})$ as above, we can write:}
\beq
    b^{\rm bias}_\alpha &=&-\frac{1}{3}\frac{\partial\zeta^{i_1i_2i_3}}{\partial b_\alpha}\left(\Si_{i_1k_1}[\Si\A]_{i_2j_2}[\Si\A]_{i_3j_3}\Ai_{j_2k_2}\Ai_{j_3k_3}\N^{(3),k_1k_2k_3}\right)^*\\\nonumber
    &&\resub{\,+\frac{1}{2}\frac{\partial\zeta^{i_1i_2i_3}}{\partial b_\alpha}\left([\Si d]_{i_1}\Si_{i_2k_2}[\Si\A]_{i_3j_3}\Ai_{j_3k_3}\N^{(2),k_2k_3}\right)^*}\\\nonumber
    &=&-\frac{1}{3}\frac{\partial\zeta^{i_1i_2i_3}}{\partial b_\alpha}\left(\Si_{i_1k_1}\av{[\Si a^{(1)}]_{i_2}[\Ai a^{(1)*}]_{k_2}[\Si a^{(2)}]_{i_3}[\Ai a^{(2)*}]_{k_3}}_{a^{(1,2)}}\N^{(3),k_1k_2k_3}\right)^*\\\nonumber
    &&\resub{\,+\frac{1}{2}\frac{\partial\zeta^{i_1i_2i_3}}{\partial b_\alpha}\left(\Si_{i_1k_1}\av{[\Si a]_{i_2}[\Ai a^*]_{k_2}\N^{(2),k_1k_2}[\Si d]_{i_3}}_a\right)^*.}
\eeq
Inserting the definition of $\beta$, we find
\beq
    b^{\rm bias}_\alpha &=&-\frac{1}{3}\av{\beta_\alpha[\Si\N^{(3)}[\Ai a^{(1)*},\Ai a^{(2)*}],\Si a^{(1)},\Si a^{(2)}]}_{a^{(1,2)}}+\resub{\frac{1}{2}\av{\beta_\alpha[\Si \N^{(2)}\Ai a^*,\Si a,\Si d]}_{a}}
\eeq
\resub{where $\N^{(3)}[u,v](\vx) = n_3(\vx)u(\vx)v(\vx)$ and $[N^{(2)}u](\vx) = d_2(\vx)u(\vx)$ up to pixel-window effects}. Since we have an efficient method of computing $\beta_\alpha$ \eqref{eq: beta-def}, this can be efficiently computed as a Monte Carlo average given some set of $N_{\rm mc}$ GRF pairs $a^{(1,2)}$, and has the same scalings as $b^{\rm num}$. 


\subsubsection{Fisher Matrix}
\noindent Computation of the Fisher matrix proceeds similarly to the shot-noise piece. Starting from \eqref{eq: bk-estimator-def}, we insert factors of the identity $\mathsf{I}\equiv \Ai \A$ via the following replacement 
\beq
    [\Si \P]_{jj'}[\Si \P]_{kk'} \to \frac{1}{2}[\Si \P]_{jl}[\Si \P]_{km}\left[\A_{ll'}\A_{mm'}+\A_{lm'}\A_{ml'}\right]\Ai_{l'j'}\Ai_{m'k'},
\eeq
effectively implementing a higher-order version of the Girard-Hutchinson estimator \citep{girard89,hutchinson90} (following \citep{2011MNRAS.417....2S,Philcox:2023psd}). Introducing two sets of iid maps $a^{(1,2)}$ with symmetric and invertible covariance $\A$, this becomes
\beq
    [\Si \P]_{jj'}[\Si \P]_{kk'} &\to& \frac{1}{4}[\Si \P]_{jl}[\Si \P]_{km}\Ai_{l'j'}\Ai_{m'k'}\\\nonumber
    &&\,\times\,\av{a^{(1)}_{l}a^{(1)}_{m}a^{(1)}_{l'}a^{(1)}_{m'}+a^{(2)}_{l}a^{(2)}_{m}a^{(2)}_{l'}a^{(2)}_{m'}-a^{(1)}_{l}a^{(1)}_{m}a^{(2)}_{l'}a^{(2)}_{m'}-a^{(2)}_{l}a^{(2)}_{m}a^{(1)}_{l'}a^{(1)}_{m'}}_{a^{(1,2)}}
\eeq
(assuming real $a$ wlog), with the particular ordering allowing most efficient use of random fields \citep{2011MNRAS.417....2S}. As such, the Fisher matrix can be written as a Monte Carlo average
\beq\label{eq: bk-fish-def}
    \textcolor{red}{\boxed{\F_{\alpha\beta}=\frac{1}{24}\av{\left(\mathsf{Q}^\dag_\alpha[\Si \P a^{(1)}]-\mathsf{Q}^\dag_\alpha[\Si \P a^{(2)}]\right)\cdot(\Si \P)\cdot\left(\mathsf{Q}_\beta[\Ai a^{(1)}]-\mathsf{Q}_\beta[\Ai a^{(2)}]\right)}^*_{a^{(1,2)}},}}
\eeq
where we have defined $\mathsf{Q}^i_\alpha[u] \equiv \partial_\alpha\zeta^{ijk}u^*_ju^*_k$. This can be computed using $N_{\rm mc}$ Monte Carlo simulations, each involving an outer product of the $N_{\rm bins}$ $\mathsf{Q}_\alpha$ vectors. By replacing two dimensions of the trilinear product by stochastic averages, we have thus converted the Fisher matrix computation to a much simpler set of matrix-vector products. Typically, convergence (relative to the statistical error) is achieved in just a few Monte Carlo iterations.

To recast $\mathsf{Q}_\alpha$ in a practically computable format, we insert the $\partial_\alpha\zeta$ definition from \eqref{eq: zeta-deriv} and \eqref{eq: Bk-Yam}:
\beq
    \mathsf{Q}_\alpha[u](\vx_1) &=& \frac{1}{\Delta_\alpha}\int_{\vk_{123}=\vec 0}{\rm d}\vx_2{\rm d}\vx_3\,e^{i\vk_1\cdot\vx_1}[e^{i\vk_2\cdot\vx_2}u^*(\vx_2)][e^{i\vk_3\cdot\vx_3}u^*(\vx_3)]\\\nonumber
    &&\,\times\,\,\left[\Theta_{b_1}(k_1)\Theta_{b_2}(k_2)\Theta_{b_3}(k_3)L_\ell(\hk_3\cdot\hx_3)+\text{5 perms.}\right].
\eeq
Replacing the Dirac delta condition by an integral as before and simplifying, we find
\beq
    \mathsf{Q}_\alpha[u](\vx) &=& \frac{2}{\Delta_\alpha}\sum_m\bar{Y}_{\ell m}(\hx)\int_{\vk}\Theta_{b_3}(k)e^{i\vk\cdot\vx}\bar{Y}_{\ell m}(\hk)\int {\rm d}\vr\,e^{i\vk\cdot\vr}g_{b_1,0}[u](\vr)g_{b_2,0}[u](\vr)\\\nonumber
    &&\,+\,\frac{2}{\Delta_\alpha}\int_{\vk}e^{i\vk\cdot\vx}\Theta_{b_1}(k)\int {\rm d}\vr\,e^{i\vk\cdot\vr}g_{b_2,0}[u](\vr)g_{b_3,\ell}[u](\vr)\\\nonumber
    &&\,+\,\frac{2}{\Delta_\alpha}\int_{\vk}e^{i\vk\cdot\vx}\Theta_{b_2}(k)\int {\rm d}\vr\,e^{i\vk\cdot\vr}g_{b_1,0}[u](\vr)g_{b_3,\ell}[u](\vr)
\eeq
with $g_{b,\ell}$ defined as in \eqref{eq: g-ell-def}. This can be directly computed using FFTs:
\begin{empheq}[box=\widefbox]{align}\label{eq: Q-bis-local}
    \mathsf{Q}_\alpha[u](\vx) &= \frac{2}{\Delta_\alpha}\sum_m\bar{Y}_{\ell m}(\hx)\mathrm{IFT}\left[\Theta_{b_3}\bar{Y}_{\ell m}\mathrm{FT}[g^*_{b_1,0}[u]g^*_{b_2,0}[u]]^*\right](\vx)\\\nonumber
    &\,+\,\frac{2}{\Delta_\alpha}\mathrm{IFT}\left[\Theta_{b_1}\mathrm{FT}[g^*_{b_2,0}[u]g^*_{b_3,\ell}[u]]^*+\Theta_{b_2}\mathrm{FT}[g^*_{b_1,0}[u]g^*_{b_3,\ell}[u]]^*\right](\vx),
\end{empheq}
requiring up to $(2\ell+5)$ FFTs, depending on symmetries. In the distant observer limit, this simplifies to 
\beq\label{eq: Q-bis-global}
    \left.\mathsf{Q}_\alpha[u](\vx)\right|_{\rm global} &=& \frac{2}{\Delta_\alpha}\mathrm{IFT}\left[\Theta_{b_3}L_\ell\mathrm{FT}[g^*_{b_1,0}[u]g^*_{b_2,0}[u]]^*\right](\vx)\\\nonumber
    &&\,+\,\frac{2}{\Delta_\alpha}\mathrm{IFT}\left[\Theta_{b_1}\mathrm{FT}[g^*_{b_2,0}[u]g^*_{b_3,\ell}[u]]^*+\Theta_{b_2}\mathrm{FT}[g^*_{b_1,0}[u]g^*_{b_3,\ell}[u]]^*\right](\vx),
\eeq
using only $4$ FFTs. Furthermore, when computing all $N_{\rm bins}$ $\mathsf{Q}_\alpha$ maps, one can carefully order \resub{the} calculation \resub{such that} each $g_{b,\ell}$ pair \resub{is Fourier-transformed} only once, requiring $(\ell_{\rm max}/2+1)N_k(N_k+1)/2$ forward FFTs in total. Moreover, one can absorb the permutation symmetries of one of the two copies of \eqref{eq: Q-bis-local} (or \eqref{eq: Q-bis-global}) as for the power spectrum, resulting in significantly faster computuation. In full, evaluation of $\F_{\alpha\beta}$ will scale with the number of bispectrum elements, $N_{\rm bins}$, due to the inverse FFTs in $\mathsf{Q}_\alpha$ (needed to apply the mask in \ref{eq: bk-fish-def}). 

In full, the Fisher matrix can be computed using the same algorithm as in \S\ref{subsubsec: pk-fish-computation}, with the only difference being that one now generates $N_{\rm mc}\lesssim 10$ \textit{pairs} of random fields $a^{(1,2)}$. As for the power spectrum, computation of the Fisher matrix is usually the rate-limiting step, though we note that (a) it scales at most linearly with the number of bins (with approximate complexity $\mathcal{O}(N_{\rm bins}N_{\rm mc}N_{\rm pix}\log N_{\rm pix})$, depending on which operations dominate, as well as $(2+N_{\rm bins})N_{\rm mc}$ applications of $\Si$), and (b) does not depend on the data, so can be efficiently precomputed. \resub{Unlike for the power spectrum, efficient computation of \eqref{eq: bk-fish-def} requires holding an array of $\Si\P\mathsf{Q}_\beta[\Ai a]$ maps in memory (of size $N_{\rm bins}$). If this resource is limiting, one can analyze a single pair of $\mathsf{Q}_{\alpha}, \mathsf{Q}_\beta$ fields at a time}: this trades off runtime for computational resources.

\subsection{Ideal Limits}\label{subsec: bk-ideal}
\noindent In the preceding sections, we have derived quasi-optimal bispectrum estimators that can be implemented entirely using FFTs and Monte Carlo summation, avoiding the $\mathcal{O}(N_{\rm pix}^3)$ operations in the na\"ive unbaised estimators. Below, we demonstrate that these expressions reduce to well-known forms in the limit of an ideal geometry (particularly applicable to simulations) or the often-used FKP weighting.

\subsubsection{Uniform Density}
\noindent As for the power spectrum, the estimators simplify significantly in the ideal limit of trivial mask ($n(\vx) = \bar{n}$) and the distant-observer approximation ($\hx,\hy=\hn$). Assuming \resub{$S(\vk) = m(\vk)\bar{n}P_{\rm fid}(\vk)$} as before, we can write the bispectrum numerator as
\beq\label{eq: bk-num-ideal}
    \left.b^{\rm num}_\alpha\right|_{\rm ideal}&=& \frac{1}{\Delta_\alpha}\frac{1}{\bar{n}^3}\int_{\vk_{123}=\vec 0}\Theta_{b_1}(k_1)\Theta_{b_2}(k_2)\Theta_{b_3}(k_3)\resub{\frac{1}{m(\vk_1)m(\vk_2)m(\vk_3)}}\frac{1}{P_{\rm fid}(\vk_1)P_{\rm fid}(\vk_2)P_{\rm fid}(\vk_3)}L_\ell(\hk_3\cdot\hn)\\\nonumber
    &&\,\qquad\qquad\times\,d^*(\vk_1)\bigg(d^*(\vk_2)d^*(\vk_3)-3\av{d^*(\vk_2)d^*(\vk_3)}\bigg),
\eeq
starting from \eqref{eq: beta-def1} and absorbing the permutations \resub{(for real $m(\vk),P_{\rm fid}(\vk)$)}. This is analogous to standard forms \citep[e.g.,][]{2012PhRvD..86f3511F}, and involves the Wiener-filtered data. Notably, \resub{$\av{d^*(\vk_2)d^*(\vk_3)} = m^2(\vk_2)\bar{n}^2\left[P(\vk_2)+1/\bar{n}\right]\delD{\vk_2+\vk_3}$}; since $d(\vk=\vec 0)=0$ (since $d$ is \resub{mean-zero}), momentum conservation forces the linear term to vanish. In general, this term contributes is non-zero in the presence of a mask or with wide-angle effects (which break translation invariance). The estimator numerator can be practically computed as before:
\beq
    &&\textcolor{blue}{\boxed{\left.b^{\rm num}_\alpha\right|_{\rm ideal}= \frac{1}{\Delta_\alpha}\int {\rm d}\vr\,g_{b_1,0}^{\rm ideal}[\Si d](\vr)g_{b_2,0}^{\rm ideal}[\Si d](\vr)g_{b_3,\ell}^{\rm ideal}[\Si d](\vr)}}\\\nonumber
    &&g_{b,\ell}^{\rm ideal}[\Si d](\vr) \equiv \frac{1}{\bar{n}}\int_{\vk}e^{i\vk\cdot\vr}\Theta_b(k)L_\ell(\hk\cdot\hn)\frac{d^*(\vk)}{P_{\rm fid}(\vk)},
\eeq
thus the numerator involves $N_{k}(\ell_{\rm max}/2+1)+1$ FFTs in total.

Via a similar argument, the shot-noise piece has the ideal limit
\beq\label{eq: bk-num-ideal}
    \left.b^{\rm bias}_\alpha\right|_{\rm ideal}&=& \frac{1}{\Delta_\alpha}\frac{1}{\bar{n}}\int_{\vk_{123}=\vec 0}\Theta_{b_1}(k_1)\Theta_{b_2}(k_2)\Theta_{b_3}(k_3)\frac{1}{P_{\rm fid}(\vk_1)P_{\rm fid}(\vk_2)P_{\rm fid}(\vk_3)}L_\ell(\hk_3\cdot\hn)\resub{\left[1+\left(\bar{n}P(\vk_1)+\text{2 perms.}\right)\right]},
\eeq
involving the true power spectrum $P(\vk)$. This is analogous to the power spectrum result \eqref{eq: pk-limit}. If $P_{\rm fid}$ is isotropic, this contributes only to the monopole (else it contributes to also to higher-order terms, though this leakage is undone by the Fisher matrix). Finally, the Fisher matrix has the limiting form
\beq\label{eq: bk-fish-ideal}
    \left.\F_{\alpha\beta}\right|_{\rm ideal} &=& \frac{1}{\Delta_\alpha\Delta_\beta}\int_{\vk_{123}=\vec 0}\Theta_{b_1}(k_1)\Theta_{b_2}(k_2)\Theta_{b_3}(k_3)L_\ell(\hk_3\cdot\hn)\frac{1}{P_{\rm fid}(\vk_1)P_{\rm fid}(\vk_2)P_{\rm fid}(\vk_3)}\\\nonumber
    &&\,\times\,\left[\Theta_{b_1'}(k_1)\Theta_{b_2'}(k_2)\Theta_{b_3'}(k_3)L_{\ell'}(\hk_3\cdot\hn)+\text{5 perms.}\right],
\eeq
for $\beta\equiv\{b_1',b_2',b_3',\ell'\}$, symmetrizing over six permutations (for even $\ell$). Notably, this vanishes unless $\{b_1,b_2,b_3\}=\{b_1',b_2',b_3'\}$ (noting that $b_1\leq b_2\leq b_3$), \textit{i.e.}\ there is no leakage between $k$-bins in the ideal limit. Furthermore, if $P_{\rm fid}(\vk)$ is isotropic and we ignore discreteness effects (such that $\int {\rm d}\hk\,L_\ell(\hk\cdot\hn)L_{\ell'}(\hk\cdot\hn) \propto \delta^{\rm K}_{\ell\ell'}$), we find
\beq\label{eq: bk-fish-ideal2}
    \textcolor{blue}{\boxed{\left.\F_{\alpha\beta}\right|_{\rm ideal}\to \delta^{\rm K}_{\alpha\beta}\frac{1}{\Delta_\alpha}\frac{N_0(\vec b)}{2\ell+1}}}
\eeq
using the definition of $\Delta_\alpha$ \eqref{eq: Delta-alpha-def}. In this limit, the Fisher matrix is diagonal, and the full estimator can be written
\beq
    \left.\hat{b}_\alpha\right|_{\rm ideal}&=& \frac{2\ell+1}{N_0(\vec b)}\frac{1}{\bar{n}^3}\int_{\vk_{123}=\vec 0}\Theta_{b_1}(k_1)\Theta_{b_2}(k_2)\Theta_{b_3}(k_3)L_\ell(\hk_3\cdot\hn)\frac{1}{P_{\rm fid}(k_1)P_{\rm fid}(k_2)P_{\rm fid}(k_3)}\frac{d^*(\vk_1)d^*(\vk_2)d^*(\vk_3)}{m(\vk_1)m(\vk_2)m(\vk_3)},
\eeq
matching the form given in \eqref{eq: ideal-bk}, which is assumed by most simulation bispectrum codes \citep[e.g.,][]{2012PhRvD..86f3511F,Foreman:2019ahr}.

Finally, we note that the normalization factor $N_\ell(\vec b)$ of \eqref{eq: N-ell-def} can be computed analytically in ideal limits (ignoring discreteness effects). Assuming an isotropic weighting, we can write
\beq
    N_\ell(\vec b) = 4\pi\int_0^\infty r^2dr\,f_{b_1,0}(r)f_{b_2,\ell}(r)f_{b_3,\ell}(r), \qquad f_{b,\ell}(r) = \int_0^\infty\frac{k_2^2dk_2}{2\pi^2}j_\ell(kr)\frac{\Theta_{b}(k)}{P_{\rm fid}(k)},
\eeq
which is evaluable with only one-dimensional numerical integrals.

\subsubsection{FKP Weights}
\noindent As for the power spectrum, the general estimators simplify significantly if we adopt the FKP weights, $\Si(\vx,\vy) = w_{\rm FKP}(\vx)\delta_{\rm D}(\vx-\vy)$. Assuming the distant observer limit for simplicity \resub{and absorbing the FKP weight into the dataset, such that $w_{\rm FKP}=1$}, the estimator numerator is given by
\beq
    \left.b^{\rm num}_\alpha\right|_{\rm FKP}&=& \frac{1}{\Delta_\alpha}\int_{\vk_{123}=\vec 0}\Theta_{b_1}(k_1)\Theta_{b_2}(k_2)\Theta_{b_3}(k_3)L_\ell(\hk_3\cdot\hn)\\\nonumber
    &&\,\times\,[\mathsf{M}^{-1}d]^*(\vk_1)\bigg([\mathsf{M}^{-1}d]^*(\vk_2)[\mathsf{M}^{-1}d]^*(\vk_3)-\left[\av{[\mathsf{M}^{-1}d]^*(\vk_2)[\mathsf{M}^{-1}d]^*(\vk_3)}+\text{2 perms.}\right]\bigg).
\eeq
This can be separably computed as before, using the filtered maps
\beq
    \left.g_{b,\ell}[\Si d](\vr)\right|_{\rm FKP} &=& \int_{\vk}e^{i\vk\cdot\vr}\Theta_b(k)L_\ell(\hk\cdot\hn)[\mathsf{M}^{-1}d]^*(\vk).
\eeq
If $P(\vk)$ is approximately constant (which is the limit used to define the FKP weights), the one-field term simplifies (and can be computed without Monte Carlo methods), though this assumption rarely holds in practice. Assuming \resub{the distant-observer limit}, the \resub{cubic} shot-noise piece can be written
\beq
    \left.b^{\rm bias}_\alpha\right|^{\rm cubic}_{\rm FKP}&=& \frac{1}{\Delta_\alpha}\int_{\vk_{123}=\vec 0}\Theta_{b_1}(k_1)\Theta_{b_2}(k_2)\Theta_{b_3}(k_3)L_\ell(\hk_3\cdot\hn)\int {\rm d}\vx\,(1-\alpha^3/\alpha_3)\bar{n}_3(\vx);
\eeq
\resub{whilst the total shot-noise is
\beq
    \left.b^{\rm bias}_\alpha\right|_{\rm FKP}&=& -2\frac{1}{\Delta_\alpha}\int_{\vk_{123}=\vec 0}\Theta_{b_1}(k_1)\Theta_{b_2}(k_2)\Theta_{b_3}(k_3)L_\ell(\hk_3\cdot\hn)\int {\rm d}\vx\,n_3(\vx)\\\nonumber
    &&\,+\, \left\{\frac{1}{\Delta_\alpha}\int_{\vk_{123}=\vec 0}\Theta_{b_1}(k_1)\Theta_{b_2}(k_2)\Theta_{b_3}(k_3)L_\ell(\hk_3\cdot\hn)d(\vk_1)d_2^*(\vk_1)+\text{2 perms.},\right\}
\eeq
involving the triply-weighted mask and doubly-weighted data defined in \eqref{eq: n3-def}. For $\ell=0$, these} forms match those used in previous (windowed) bispectrum analyses \citep[e.g.,][]{2015MNRAS.451..539G,2017MNRAS.465.1757G}.

Due to the non-trivial mask, the Fisher matrix has a more complex limiting form:
\beq
    \F_{\alpha\beta} &=& \frac{1}{\Delta_\alpha\Delta_\beta}\int_{\vk_{123}=\vec 0}\int_{\vk_{123}'=\vec 0}n^*(\vk_1+\vk_1')n^*(\vk_2+\vk_2')n^*(\vk_3+\vk_3')\\\nonumber
    &&\,\times\,\Theta_{b_1}(k_1)\Theta_{b_2}(k_2)\Theta_{b_3}(k_3)L_\ell(\hk_3\cdot\hn)\left[\Theta_{b_1'}(k_1')\Theta_{b_2'}(k_2')\Theta_{b_3'}(k_3')L_{\ell'}(\hk_3'\cdot\hn)+\text{5 perms.}\right],
\eeq
similar to the power spectrum result \eqref{eq: pk-fkp-lim}. This essentially involves the overlap of two $\vk$-space volumes, modulated by the window function of the (FKP-weighted) mask, and straightforwardly relaxes to the ideal limit if $n$ is uniform (whence $\vk_i+\vk_i'\to\vec 0$). This differs from the normalization found in conventional windowed bispectrum estimators, which, in our notation, can be written:
\beq
    \F_{\alpha\beta}^{\rm conv} &=& \frac{1}{\Delta_\alpha\Delta_\beta}\int_{\vk_{123}=\vec 0}\Theta_{b_1}(k_1)\Theta_{b_2}(k_2)\Theta_{b_3}(k_3)L_\ell(\hk_3\cdot\hn)\left[\Theta_{b_1'}(k_1)\Theta_{b_2'}(k_2)\Theta_{b_3'}(k_3)L_{\ell'}(\hk_3\cdot\hn)+\text{5 perms.}\right]\\\nonumber
    &&\,\times\,\int {\rm d}\vx\,n^3(\vx)\\\nonumber
    &=& \frac{\delta_{\alpha\beta}^{\rm K}}{\Delta_\alpha}\frac{N_0(\vec b)}{2\ell+1}\int {\rm d}\vx\,n^3(\vx),
\eeq
for bin-volume $N_0(\vec b)$ defined in \eqref{eq: N-ell-def} (with $P_{\rm fid}(k)=1$). In this limit, the full estimator (dropping the linear and bias terms) becomes
\beq
    \left.\hat{b}_\alpha^{\rm conv}\right|_{\rm FKP} = \frac{2\ell+1}{N_0(\vec b)}\left[\int {\rm d}\vx\,n^3(\vx)\right]^{-1}\int {\rm d}\vr\,g_{b_1,0}^{\rm ideal}[\mathsf{M}^{-1}d](\vr)g_{b_2,0}^{\rm ideal}[\mathsf{M}^{-1}d](\vr)g_{b_3,\ell}^{\rm ideal}[\mathsf{M}^{-1}d](\vr),
\eeq
matching standard forms \citep[e.g.,][]{2015MNRAS.451..539G,2017MNRAS.465.1757G}. As for the power spectrum, the difference between the two estimators occurs since our approach estimates unwindowed bispectra, whilst the traditional approach computes window-convolved bispectra, which must be compared to similarly convolved theory models.

\section{Code Implementation}\label{sec: implementation}

\noindent In \polybin\footnote{\href{https://github.com/oliverphilcox/PolyBin3D}{GitHub.com/OliverPhilcox/PolyBin3D}} we provide a \textsc{Python} implementation of the power spectrum and bispectrum estimators discussed above, mirroring that of the \textsc{PolySpec} code \citep{PolyBin,Philcox4pt2} (which implements analogous estimators for fields on the two-sphere). For each statistic, we provide routines for computing two types of estimator: `\textcolor{red}{unwindowed}' and `\textcolor{blue}{ideal}', respectively implementing the main estimators of this work (the \textcolor{red}{red} and \textcolor{blue}{blue} equations in the above), and the simplified forms for uniform $n(\vx)$, as appropriate for periodic-box simulations. This code makes extensive use of the \textsc{fftw}, \textsc{mkl\_fft}, or \textsc{jax} packages\footnote{Available at \href{https://github.com/pyFFTW/pyFFTW}{github.com/pyFFTW/pyFFTW}, \href{https://github.com/IntelPython/mkl_fft}{github.com/IntelPython/mkl\_fft} and \href{https://docs.jax.dev/en/latest/}{docs.jax.dev}.} \citep{fftw} to perform FFTs. \resub{Rate limiting steps of the computation (including summations and multiplications in Fourier- and real-space) are computed in \textsc{c} using parallelized \textsc{cython} code.} The code computes the estimators for arbitrary (user-defined) weighting scheme $\Si$ and mask $n(\vx)$, and is written to \resub{use as few FFTs as possible}.

The code additionally supports GPU acceleration through \textsc{jax}, \resub{which provides a significant speed boost (up to $\sim 5\times$ in typical use-cases) with minimal addition of code.} Since the computation of \resub{bispectrum} can be memory intensive, especially at higher grid sizes and/or fine binning, special care has been taken to manage GPU memory usage \resub{-- that said, we caution that the GPU code can suffer from memory exhaustion when analyzing high-resolution datasets}. 
\resub{In general, we find most significant acceleration when computing the power spectrum, with a single GPU outperforming a $64$-core CPU node by up to $10\times$. When computing the bispectrum Fisher matrix, we do not find enhanced performance due to the large number of transfers of data between the CPU and GPU required to avoid saturating the memory of the latter.}

The power spectrum module, \texttt{PSpec}, contains the code required to compute the bandpowers of observational datasets and simulation boxes. This contains the following main routines:
\begin{itemize}
    \item \texttt{\textcolor{red}{Pk\_numerator}}: This computes the power spectrum numerator given in \eqref{eq: pk-num-fft}, optionally adopting the distant-observer approximation.
    \item \texttt{\textcolor{red}{compute\_fisher\_contribution}, \textcolor{red}{compute\_shot\_contribution}}: These compute the contributions to the Fisher matrix and shot-noise numerator from one of $N_{\rm mc}$ random maps, implementing \eqref{eq: fish-from-Q}, with $\mathsf{Q}_\alpha$ maps computed from \eqref{eq: Q-alpha-local} or \eqref{eq: Q-alpha-global} depending on the choice of line-of-sight. 
    \item \texttt{\textcolor{red}{compute\_theory\_contribution}}: This computes the contribution to the correction matrix $\mathcal{G}_{\alpha\iota}$ given in \eqref{eq: theory-matrix-def}, which relates the coarsely binned statistic to the finely binned input theory, as detailed in Appendix \ref{app: binning-theory}. This is computed analogously to the Fisher matrix, and can be optionally included if one wishes to bin-integrate the theoretical model. 
    \item \texttt{\textcolor{red}{compute\_covariance\_contribution}}: \resub{This computes the contribution to the Gaussian covariance matrix of $p^{\rm num}_\alpha$, given some input theoretical power spectra. This proceeds analogously to the Fisher matrix, with the replacement $\Si\P\to\Si\C\Sit\equiv \Si[\P\xi\P^\dagger+\N]\Sit$.}
    \item \texttt{\textcolor{red}{Pk\_unwindowed}}: This wraps the \texttt{Pk\_numerator} routine to compute the full unwindowed power spectrum estimator on a specified dataset, given the Fisher matrix and shot-noise.
    \item \texttt{\textcolor{blue}{Pk\_numerator\_ideal}}: This computes the idealized power spectrum numerator given in \eqref{eq: pk-limit}, optionally adopting the distant-observer approximation. Data are weighted by a fiducial power spectrum monopole, $P_{\rm fid}(k)$.
    \item \texttt{\textcolor{blue}{compute\_fisher\_ideal}}: This computes the idealized Fisher matrix given in \eqref{eq: pk-limit}, assuming the $S(\vk) = m(\vk)\bar{n}P_{\rm fid}(k)$ weighting scheme and assuming a spatially-constant line-of-sight. We can optionally ignore discreteness effects, such that the Fisher matrix \resub{agrees with the idealized limit} \eqref{eq: fish-ideal-continuous}. 
    \item \texttt{\textcolor{blue}{Pk\_ideal}}: This wraps the above two routines to compute the idealized power spectrum estimator on a specified dataset.
\end{itemize}
Similarly, the bispectrum module, \texttt{BSpec}, computes the binned bispectrum coefficients, and comprises similar routines:
\begin{itemize}
    \item \texttt{\textcolor{red}{Bk\_numerator}}: This computes the bispectrum numerator given in \eqref{eq: b-numerator-practical} using the $\beta$ definition of \eqref{eq: beta-def}, optionally adopting the distant-observer approximation. By default, the code computes only the cubic term in \eqref{eq: b-numerator-practical}; however, the linear term can also be included using suite of externally defined \resub{simulations} (\textit{i.e.}\ mock catalogs) or internally generated Gaussian random fields.
    \item \texttt{\textcolor{red}{compute\_fisher\_contribution}, \textcolor{red}{compute\_shot\_contribution}}: This computes the contribution to the Fisher matrix or shot-noise numerator \resub{from a pair of random maps}, implementing \eqref{eq: bk-fish-def}, with $\mathsf{Q}_\alpha$ maps computed from \eqref{eq: Q-bis-local} or \eqref{eq: Q-bis-global} depending on the choice of line-of-sight. 
    \item \texttt{\textcolor{red}{compute\_covariance\_contribution}}: \resub{This computes the contribution to the Gaussian covariance matrix of $\widehat{b}^{\rm num}_\alpha$, as for the power spectrum (including mask-induced effects).}
    \item \texttt{\textcolor{red}{Bk\_unwindowed}}: This wraps the \texttt{Bk\_numerator} routine to compute the full unwindowed bispectrum estimator on a specified dataset, given the Fisher matrix.
    \item \texttt{\textcolor{blue}{Bk\_numerator\_ideal}}: This computes the idealized bispectrum numerator given in \eqref{eq: bk-num-ideal}, optionally adopting the distant-observer approximation. Data are weighted by a fiducial power spectrum monopole, $P_{\rm fid}(k)$, and we do not include the linear term in the numerator.
    \item \texttt{\textcolor{blue}{compute\_fisher\_ideal}}: This computes the idealized Fisher matrix given in \eqref{eq: bk-fish-ideal}, assuming the $S(\vk) = m(\vk)\bar{n}P_{\rm fid}(k)$ weighting scheme and assuming a spatially-constant line-of-sight. We can optionally ignore discreteness effects, such that the Fisher matrix is given by \resub{the ideal limit} \eqref{eq: bk-fish-ideal2}. 
    \item \texttt{\textcolor{blue}{Bk\_ideal}}: This wraps the above two routines to compute the idealized bispectrum estimator on a specified dataset.
\end{itemize}
We further include a number of utility functions, including \texttt{get\_ks} (to return the $k$-bins used in the estimators) and \texttt{generate\_data} (to create Gaussian random field data with a specified set of power spectrum multipoles.

\begin{figure}
    \centering
\noindent\begin{minipage}{.45\textwidth}
\begin{lstlisting}{Name}
# Import code
import PolyBin3D as pb

# Load base class specifying dimensions
base = pb.PolyBin3D(boxsize, gridsize)                    

# Load power spectrum class, specifying binning, mask and filter 
pspec = pb.PSpec(base, k_bins, lmax=lmax, mask=mask, applySinv=applySinv)

# Compute Fisher matrix
fish = pspec.compute_fisher(N_mc)

# Compute windowed power spectra
Pk_ideal = pspec.Pk_ideal(data) 

# Compute unwindowed power spectra
Pk_unwindowed = pspec.Pk_unwindowed(data, fish=fish)
\end{lstlisting}
\end{minipage}\hfill
\begin{minipage}{.45\textwidth}
\begin{lstlisting}{t}
# Import code
import PolyBin3D as pb

# Load base class specifying dimensions
base = pb.PolyBin3D(boxsize, gridsize)                    

# Load bispectrum class, specifying binning, mask and filter
bspec = pb.BSpec(base, k_bins, lmax=lmax, mask=mask, applySinv=applySinv)

# Compute Fisher matrix
fish = bspec.compute_fisher(N_mc)

# Compute windowed bispectra
Bk_ideal = bspec.Bk_ideal(data) 

# Compute unwindowed bispectra
Bk_unwindowed = bspec.Bk_unwindowed(data, fish=fish, include_linear_term=False)
\end{lstlisting}
\end{minipage}
    \caption{Sample \textsc{Python} code for computing the power spectrum (left) and bispectrum (right) of a dataset (\texttt{data}) using the \polybin package. Extensive tutorials demonstrating the various functionalities of the package are available on GitHub.}
    \label{fig: sample-code}
\end{figure}

In Fig.\,\ref{fig: sample-code}, we show sample code to compute the power spectrum and bispectrum of a three-dimensional dataset using \polybin. For the idealized estimators, one simply needs to specify the data and binning strategy (and optionally a filtering scheme and mask), whilst the unwindowed estimators require the Fisher matrix to be estimated before the data is analyzed. \resub{Depending on the data-set size, this step can be expensive and would usually be performed on a high-performance computing cluster}, though we stress that it does not depend on the data. Detailed tutorials discussing the various estimators and their implementation (including validation and application to observational datasets) can be found online.

\section{Validation}\label{sec: validation}
\noindent We now provide a practical demonstration of the above estimators and validate that they return unbiased estimates of the power spectrum and bispectrum. For this, we will construct a variety of mock-based tests, centered around a suite of Gaussian random fields with a known power spectrum (a linearly biased spectrum, with parameters similar to \citep{2020A&A...641A...6P}), gridding with a Nyquist frequency of $0.6\,\hMpc$. Furthermore, we will often add a multiplicative mask based on the BOSS LOWZ-South sample \citep{2015ApJS..219...12A} (with a volume $V\approx0.4\,h^{-3}\mathrm{Gpc}^3$), and further assume a `triangle-shaped-cell' pixel-window convolution scheme, with a global line-of-sight for anisotropies. This builds significantly upon the validation schemes of previous works \citep[e.g.,][]{2021PhRvD.103j3504P,Philcox:2021ukg}, and now includes (a) explicit validation of the unwindowed estimators when applied to masked data, (b) validation of the bispectrum estimators on simulations containing an injected three-point function. \resub{We additionally compare our power spectrum measurements to those obtained using other published codes, finding excellent agreement.}

\subsection{Power Spectrum}\label{subsec: pk-validation}
\subsubsection{Simulation Validation}\label{subsubsec: pk-valid-sub}
\noindent First, we validate the power spectrum estimators. For this, we generate $1000$ Gaussian random field simulations with known power spectrum in the range $k\in[0.02,0.40)\,\hMpc$, then use \polybin to measure the binned power spectrum multipoles across 132 bins with $k\in[0.01,0.45)\,\hMpc$ and $\ell=\{0,2,4\}$.\footnote{We fix $P(k)=0$ for $k$ outside the range of interest to remove bias from unmeasured bins; in practice, any bias can be alleviated by dropping the final few bins in the estimator} We consider three analyses: (1) the ideal estimator applied to unmasked data; (2) the ideal estimator applied to masked data; (3) the unwindowed estimator applied to masked data. For (3), the Fisher matrix is computed using a further $1000$ simulations (which can be embarrassingly parallelized). If the estimator is unbiased, we expect the power spectra of (1) and (3) to agree, whilst the differences between (1) and (2) will show the bin-convolution effects of the survey geometry on the power spectrum. In all cases, we will assume the FKP form for $\Si$ \eqref{eq: Sinv-fkp}, though we discuss the effects of optimal weights in \S\ref{subsec: optimal-weight-validation}.

\begin{figure}
    \centering
    \includegraphics[width=0.7\textwidth]{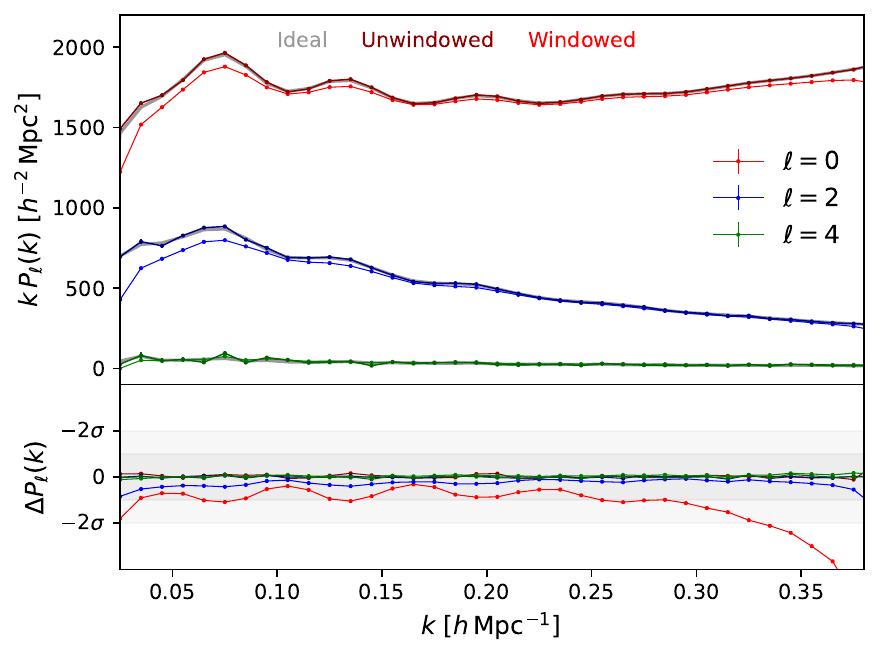}
    \caption{Validation of the \polybin power spectrum estimators. The top panel shows the mean and standard error of power spectrum bandpowers estimated from $1000$ Gaussian random simulations, with results including (not including) a realistic survey mask shown in colors (grey). When the mask is included, we give results from both the standard windowed estimators (light colors, equal to the `idealized' estimators discussed herein) and the unwindowed estimators discussed in this work (dark colors). The bottom panel shows deviations of the masked spectra compared to the ideal `truth', in terms of the error-bars of a single $0.4\,h^{-3}\mathrm{Gpc}^3$ volume. We do not subtract shot-noise contributions in any case. We find excellent agreement between the ideal and unwindowed estimators; in contrast, the mask induces strong distortions in the windowed spectra, particularly on large scales.}
    \label{fig: pk-valid}
\end{figure}

\begin{figure}
    \centering
    \includegraphics[width=0.49\textwidth]{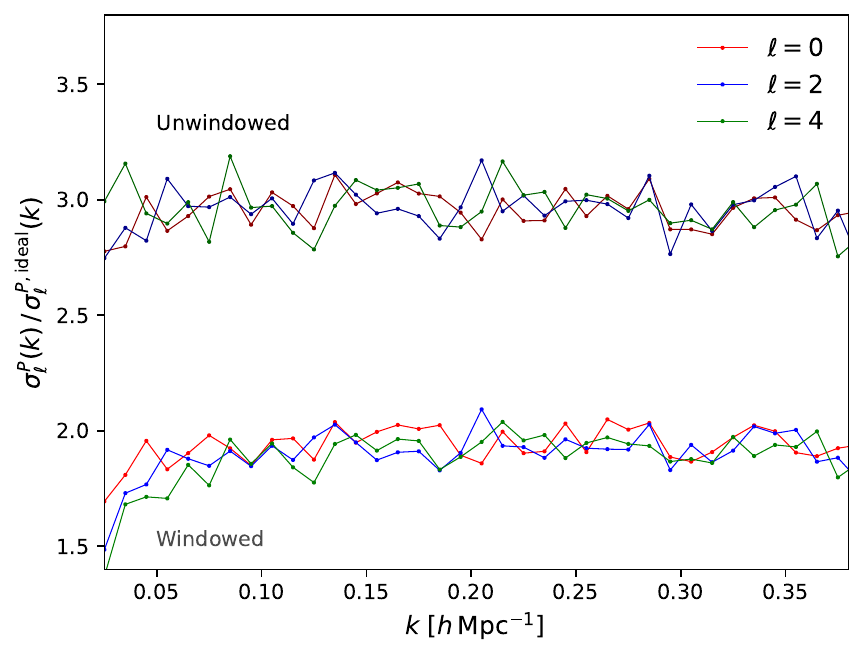}
    \includegraphics[width=0.49\textwidth]{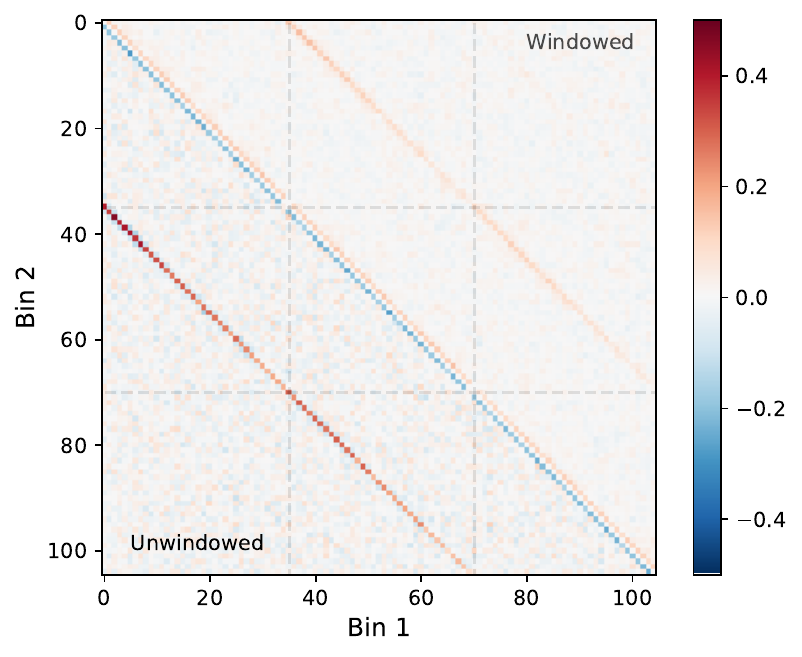}
    \caption{Comparison of the covariances from the windowed and unwindowed power spectra shown in Fig.\,\ref{fig: pk-valid}. In the left panel, we show the ratio of the errors from the masked and unmasked simulations: as expected, masking significantly inflates the errors. In the right panel, we plot the (diagonal-subtracted) correlation matrices ($\mathbb{C}_{ij}/\sqrt{\mathbb{C}_{ii}\mathbb{C}_{jj}}$ for covariance $\mathbb{C}$), with the windowed (unwindowed) results shown in the upper right (lower left) triangle. We stack results for the three Legendre multipoles, with $\ell=0$ shown in the top left panel. Though the unwindowed multipoles exhibit higher variances, they do not contain less signal-to-noise, due to the differing correlation structures, in particular the anti-correlations along the leading diagonal.}
    \label{fig: pk-errors}
\end{figure}

The resulting power spectra are shown in Fig.\,\ref{fig: pk-valid}, with the covariances displayed in Fig.\,\ref{fig: pk-errors}. Comparing the windowed and ideal spectra, we observe significant ($\sim 8\sigma$, for our $0.4\,h^{-3}\mathrm{Gpc}^3$ volume) distortions induced by the survey mask (beyond the volume rescaling, which is already accounted for), which must be taken into account in any theoretical model. These are particularly notable both on large-scales and for the quadrupole moment (and on the smallest-scales, due to shot-noise corrections). In contrast, the mean power spectra obtained from our unwindowed estimators are highly consistent with those from ideal simulations (without a mask) across all scales and multipoles, indicating that our pipeline is unbiased, as desired. As such, the unwindowed power spectrum estimates can be directly compared to theory, without need for mask-convolution of the latter.

As shown in the left panel of Fig.\,\ref{fig: pk-errors}, there are significant differences in the variances of the unwindowed and windowed power spectrum estimates. Whilst both are larger than the ideal spectra (as expected, due to the lower effective volume), the unwindowed estimates have $\approx 2\times$ larger error, roughly consistent across all bins and multipoles. This may appear somewhat alarming; if our $\Si$ weighting scheme is close-to-optimal, we expect that the unwindowed estimators should achieve approximately minimum-variance error-bars. This discrepancy can be resolved by looking at the matrix correlation structure (right panel of Fig.\,\ref{fig: pk-errors}); the unwindowed estimators have negative correlations between neighboring bins in contrast to the positive correlations for windowed estimators (and diagonal structure for the unmasked data), which results in an approximately equal signal-to-noise ($\approx 440$) from each estimator (in fact $10\%$ higher for the unwindowed approach across the non-trivial bins). In each case, we see also contributions between different Legendre moments, which are sourced both by the anisotropic clustering and the mask. One might na\"ively have expected the unwindowed estimators to have a diagonal covariance (\textit{i.e.}\ for the matrix $\F^{-1}$ to undo any mask-induced correlations); as discussed in \S\ref{sec: estimator-theory}, the action of $\F^{-1}$ is only to remove mask-induced \textit{biases}, and the covariance will almost always contain off-diagonal contributions.\footnote{In the limit of ideal weights, the covariance of the data $\hat{p}$ is given by $\mathcal{F}^{-1}$, though the covariance of $\mathcal{F}^{1/2}\hat{p}$ is indeed diagonal, where $\mathcal{F}^{1/2}$ is the Cholesky factorization of $\F$ \citep[e.g.,][]{Hamilton:2005ma}.}

\subsubsection{Timings}
\noindent Before continuing, it is useful to assess the practicalities of our estimator. For the set-up considered herein, computing the power spectrum numerators for each simulation required \resub{$\approx 0.5$ seconds on one 64-core node}, whilst the Fisher matrix \resub{and shot-noise contribution} needed for unwindowed estimators required $90$ seconds and \resub{$4$ seconds} per Monte Carlo iteration respectively, using a grid-size of \resub{$194\times 365\times 202$} (to obtain $k_{\rm Ny}=0.6\,\hMpc$ for our sample). \resub{The \textsc{jax} implementation is considerably faster: after compilation, the numerator, Fisher matrix realization and shot-noise realization require $0.03$, $6$ and $2$ seconds respectively on an A100 GPU.} There results match the number of FFTs required; each numerator requires just one FFT, whilst the Fisher matrix requires $266$, matching the scalings discussed in \S\ref{subsec: pk-implementation} (given the 132 bins in our test). Whilst the Fisher matrix is \resub{a little more expensive than the numerator}, we remind the reader that this does not depend on the data, thus only has to be computed once for a given set of bins and mask. Furthermore, the number of Monte Carlo iterations used in the above tests ($N_{\rm mc}=1000$) is conservative; reducing to $N_{\rm mc}=100$ gives a (stochastic) \resub{mean absolute} error of $\approx 0.1\sigma$, or $\approx 0.3\sigma$ with $N_{\rm mc}=25$.

\subsubsection{Code Comparison}

\begin{figure}
    \centering
    \includegraphics[width=0.49\linewidth]{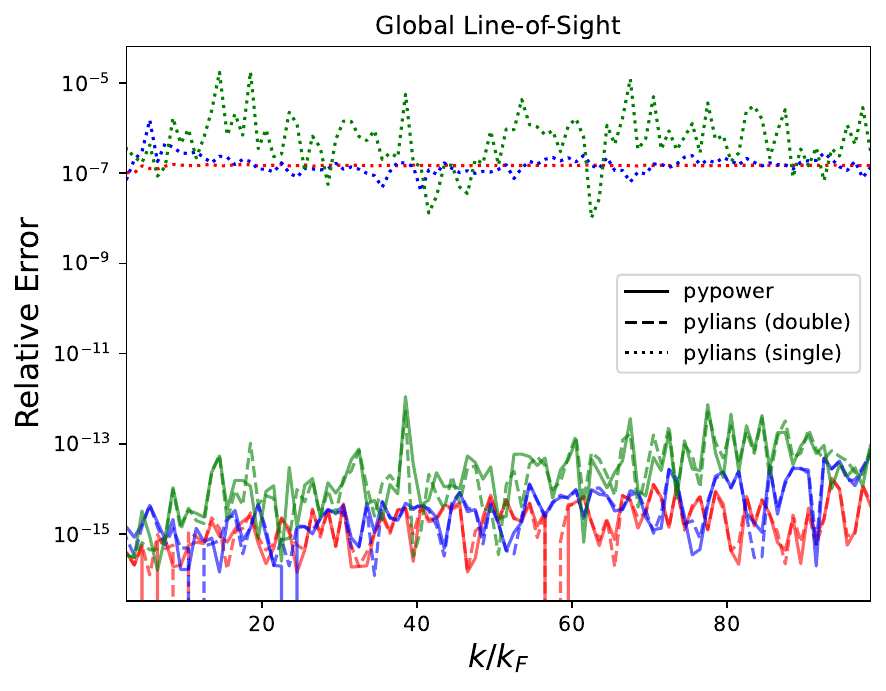}
    \includegraphics[width=0.49\linewidth]{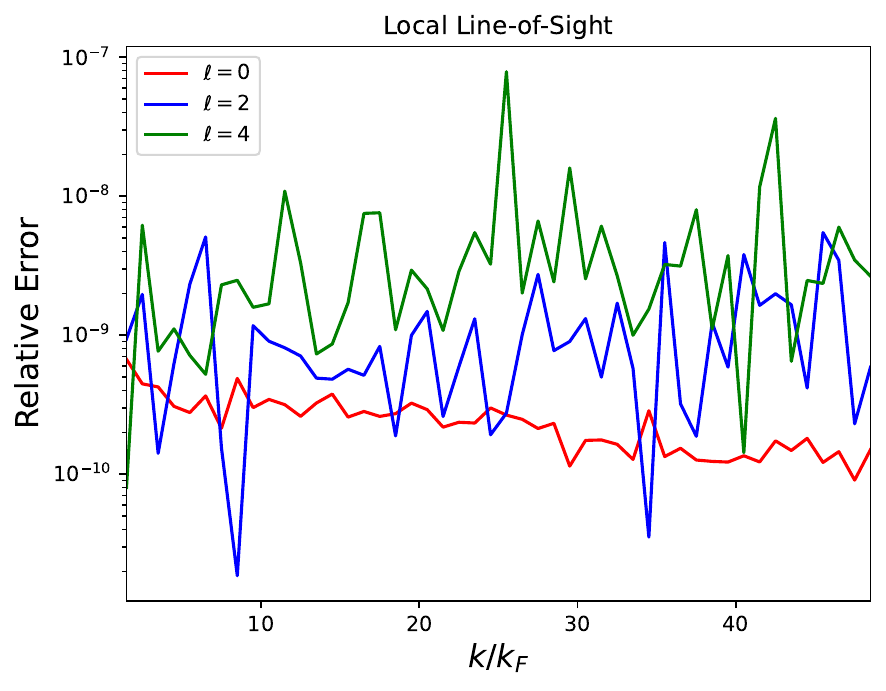}
    \caption{\resub{Comparison of the \polybin power spectrum multipoles to those from two independent codes: \textsc{PyPower} and \textsc{Pylians}. The left and right panels show results for a global and local line-of-sight respectively. For \textsc{pylians} (which includes only global line-of-sight estimators), we show results using both single- and double-precision calculations (\textsc{float32} and \textsc{float64}). In all cases, and for all multipoles, we find excellent agreement between codes with a double-precision error within $0.00001\%$.}}
    \label{fig: code-comparison}
\end{figure}
\noindent \resub{As a further test, we compare the \polybin power spectra to those obtained using other codes. To this end, we first generate a test dataset comprising a Gaussian random field with boxsize $1000\Mpch$ and $256^3$ grid cells. We then measure the power spectrum using three codes: \polybin, \textsc{PyPower} (based on \textsc{nbodykit}), and \textsc{Pylians} \citep{Pylians,2017JCAP...07..002H,2018AJ....156..160H}.}\footnote{\resub{Available at \href{https://pypower.readthedocs.io/en/latest/}{pypower.readthedocs.io} and \href{https://pylians3.readthedocs.io}{pylians3.readthedocs.io}. Note that we use \textsc{PyPower} rather than \textsc{nbodykit} since the latter code contains minor inaccuracies in the local line-of-sight estimator.}}
\resub{To ensure compatibility with all codes, we use a global line-of-sight and a cubic Fourier-space grid. We perform two slight modifications to the \textsc{Pylians} code: we add a double-precision version (instead of the native single-precision computation), and we allow for a more general $k$-space binning strategy.}
\footnote{\resub{In \textsc{Pylians}, all frequencies are measured with respect to the fundamental mode of the (cubic) box, $k_{\rm F}$.} As such, the binning operation involves only integer arithmetic, unlike the floating-point arithmetic required in other codes. By default, \textsc{Pylians} bins in integer units of $k_{\rm F}$ -- translating this to other codes leads to numerical error due to the inherent difficulties of comparing almost identical floats.} 
\resub{We additionally compare local line-of-sight power spectra between using \polybin and \textsc{PyPower}.}\footnote{\resub{For this test, we omit the pixel-space window function. This is treated differently between \polybin and \textsc{PyPower} -- \polybin deconvolves it \textit{before} applying the Legendre polynomial factor to the dataset, whilst \textsc{PyPower} (and \textsc{nbodykit}), deconvolve it at the end (\textit{i.e.}\ our algorithm computes $\int d\vx\,e^{-i\vk\cdot\vx}L_\ell(\hk\cdot\hx)[\mathsf{M}^{-1}d](\vx)$ instead of $m^{-1}(\vk)\int d\vx\,e^{-i\vk\cdot\vx}L_\ell(\hk\cdot\hx)d(\vx)$.}}

\resub{The results are shown in Fig.\,\ref{fig: code-comparison}. For the global line-of-sight, we find excellent agreement between all three codes for all scales and multipoles, with \polybin and \textsc{PyPower} consistent to within $1$ part in $10^{12}$ (consistent with machine error). For \textsc{Pylians}, we find agreement within $1$ part in $10^{5}$, or $10^{12}$ if we switch to double-precision. For the local line-of-sight, the agreement is again excellent -- in this case \textsc{PyPower} and \polybin agree to within $1$ part in $10^{7}$ (or $10^9$ for the monopole). This implies that our estimator can be robustly applied to both numerical simulations and observational data. More generally, \polybin has the following advantages compared to the other codes:
\begin{itemize}
    \item We can compute the power spectra of both cubic and non-cubic geometries, appropriate to both simulations and observational data (unlike \textsc{Pylians}, which is restricted to the former).
    \item We can take advantage of the fast GPU routines available in \textsc{jax} (which is not possible with other off-the-shelf codes, though see \textsc{BFast} for a fast \textsc{jax}-based simulation bispectrum estimator).\footnote{\resub {Available at \href{https://github.com/tsfloss/BFast}{github.com/tsfloss/BFast}.}}
    \item We can apply an arbitrary linear weighting scheme $\Si$ to the data and fully account for the corresponding modification to the shot-noise and normalization matrix. This additionally allows implementation of minimum-variance estimators. 
    \item We can analyze both the power spectrum and bispectrum in a unified framework, and directly compute the statistics, their window matrices, and their Gaussian covariances.
\end{itemize}
}

\subsubsection{Optimal Weights}\label{subsec: optimal-weight-validation}
\noindent Next, we consider the impact of the $\Si$ weighting scheme on our power spectrum measurements. For this purpose, we perform a similar analysis to \resub{that of \S\ref{subsubsec: pk-valid-sub}}, but compute power spectra using both the FKP weighting scheme \eqref{eq: Sinv-fkp} and the optimal solution specified by \eqref{eq: Sinv-optimal}. To ensure correct treatment of stochastic effects, we first generate 250 masked Gaussian random field simulations as above, but nulling the shot-noise contribution to the input power spectrum. We then add a Poisson noise contribution to each scaling as $n(\vx)$, which is then pixel-window-convolved, emulating the observational data. The optimal weights are realized by solving \eqref{eq: cgd-operation-eqn} via conjugate gradient descent using ten iterations (preconditioned on the ideal solution of \ref{eq: Sinv-ideal}).\footnote{This is sufficient to ensure convergence at the $0.1\%$ level, and cannot induce bias.} Power spectra are computed using the same binning schemes as before, and we subtract the measured shot-noise in both cases (which differs slightly in each case, due to the differing weighting schemes adopted).

\begin{figure}
    \centering
    \includegraphics[width=0.6\textwidth]{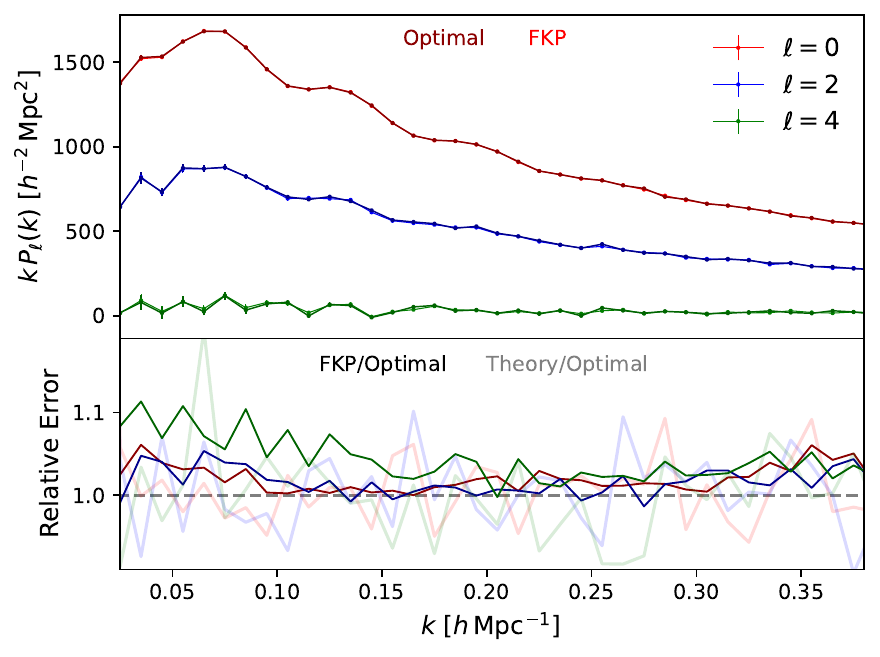}
    \caption{Comparison of shot-noise-subtracted power spectrum multipoles obtained with the unwindowed estimators of this work applied to 250 Gaussian simulations using two choices of weighting schemes: optimal (dark colors) and FKP (light colors). The bottom panel shows the ratio of the errorbars from the two weighting schemes (dark) and between the theoretical prediction and optimal weights (light). We find excellent agreement between the mean spectra, with the optimal scheme leading to $5-10\%$ improved errors, which closely match those expected.}
    \label{fig: pk-opt-errors}
\end{figure}

\begin{figure}
    \centering
    \includegraphics[width=0.49\textwidth]{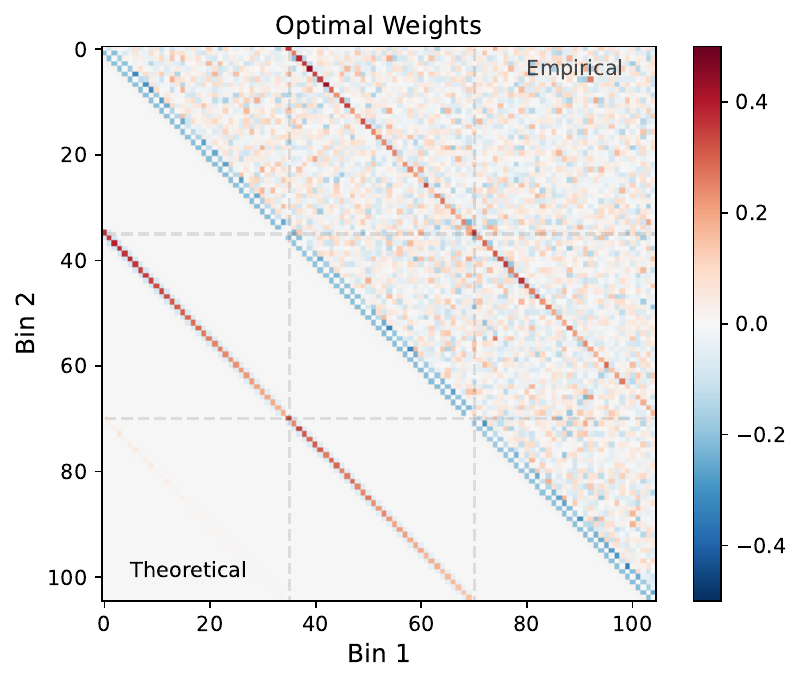}
    \includegraphics[width=0.49\textwidth]{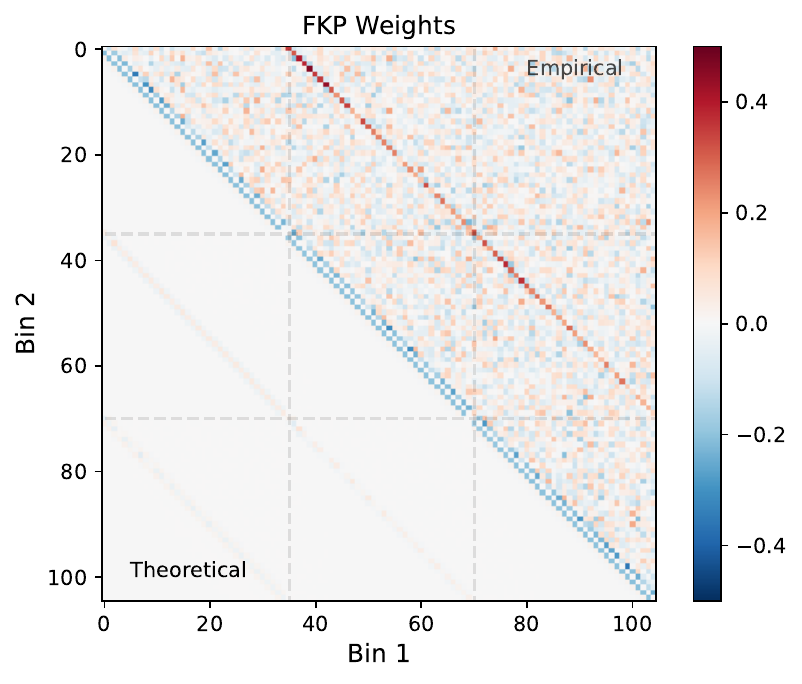}
    \caption{Correlation matrices for the power spectra shown in Fig.\,\ref{fig: pk-opt-errors}, using $250$ simulations analyzed with optimal (left) and FKP (right) weighting schemes. In each panel, we show the theoretical prediction (the inverse Fisher matrix) in the lower-left triangle. Notably, this is a good approximation of the true covariance for the optimal weighting only, implying that the optimal weights are indeed close to optimal.}
    \label{fig: pk-opt-cov}
\end{figure}

In Fig.\,\ref{fig: pk-opt-errors}, we show the resulting power spectrum multipoles. As expected, no bias is induced by changing $\Si$, however, we find moderate differences in the variances, with the FKP weights leading to $5-10\%$ inflated errors at low-$k$ (particularly for the hexadecapole). This has an important conclusion: optimal weights lead to more precise power spectrum measurements. That this predominantly affects large-scales and higher-multipoles makes sense since these components are less well approximated by the (shot-noise-dominated) FKP limit. Furthermore, the optimal errors are seen to be in good agreement with the predictions from the inverse Fisher matrix, $\F^{-1}$, suggesting that our estimators are close to minimum variance, as desired. This is further shown by the correlation structures shown in Fig.\,\ref{fig: pk-opt-cov}; $\F^{-1}$ closely matches the empirical covariance for optimal weights (including the off-diagonal components), but there are significant deviations when using FKP weights. 

Despite the slight gains in signal-to-noise benefits (which would be more pronounced for a survey with lower shot-noise), using optimal weights comes at the expense of increased computational cost. \resub{Computing the Fisher matrix and shot-noise required $36$ and $0.5$ minutes respectively per iteration on a CPU node}, with the increase due to the need to solve the optimality condition for each power spectrum bin. Furthermore, the power spectrum numerators required \resub{$20$} seconds per simulation (and 53 FFTs). \resub{When run on an A100 GPU (after just-in-time compilation), the numerator requires $0.9$ seconds, whilst a single shot-noise and Fisher matrix iteration need $4$ and $120$ seconds respectively}. In practice, computation may be expedited by using more efficient conjugate gradient descent solvers (or via less numerical iterations) or using alternative methods to compute $\Si_{\rm opt}$ \citep[e.g.,][]{Munchmeyer:2019kng}.

\subsection{Bispectra}
\subsubsection{Set-Up}
\noindent Next, we provide a numerical validation of the bispectrum estimators. For this, we adopt a similar methodology to before, but now inject a known bispectrum into the simulations. This is done by first generating a Gaussian random field, $\delta_G(\vx)$ with known power spectrum $P_G(\vk)$, then performing the redefinition \citep[e.g.,][]{2011MNRAS.417....2S}
\beq
    \delta_G(\vk)\to \delta_G(\vk)+\frac{\epsilon}{6}\beta(k)\int {\rm d}\vx\,e^{-i\vk\cdot\vx}\int_{\vk_2}\frac{\beta(k_2)}{P_G(\vk_2)}\delta_G(\vk_2)e^{i\vk_2\cdot\vx}\int_{\vk_3}\frac{\beta(k_3)}{P_G(\vk_3)}\delta_G(\vk_3)e^{i\vk_3\cdot\vx}.
\eeq
For small $\epsilon$, this produces an isotropic bispectrum $B(k_1,k_2,k_3)=\epsilon\beta(k_1)\beta(k_2)\beta(k_3)$. For definiteness, we here assume $\epsilon = 0.25$ and $\beta(k) = P^{2/3}_G(k)$ (without shot-noise), filtering all fields to the $k$-range $k\in[0.1,0.4)\,\hMpc$. Since the dimensionality of the bispectrum is much larger than the power spectrum, we adopt a coarser binning with $\delta k = 0.05\,\hMpc$ for $k\in[0.05,0.45)\,\hMpc$ and $\ell=\{0,2\}$, which corresponds to 196 total configurations. We analyze 500 simulations in total using the FKP form of $\Si$, and initially drop the linear term in the estimator.

\begin{figure}
    \centering
    \includegraphics[width=0.6\textwidth]{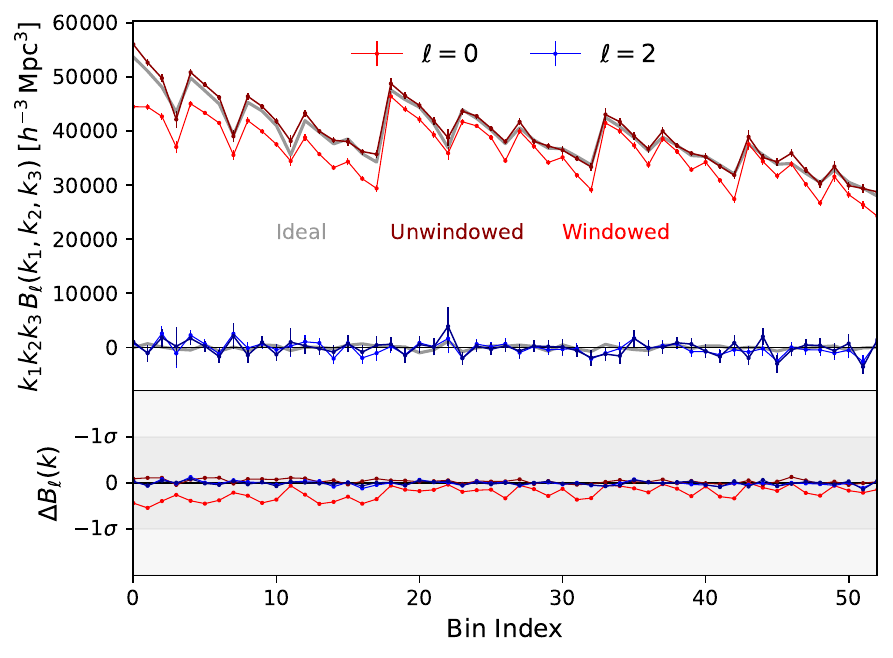}
    \caption{Validation of the \polybin bispectrum estimators. Analogous to Fig.\,\ref{fig: pk-valid}, the top panel shows the bispectrum coefficients computed from simulations without a mask applied (grey), as well as the unwindowed (dark, defined in this work) and windowed (light, conventional) estimators applied to masked data. Here, we use $500$ simulations, which each contain an injected bispectrum monopole. We collapse the three-dimensional bispectrum statistic into one dimension, ordering the $\{k_1,k_2,k_3\}$ bin triplet by first iterating over $k_3$ at $k_1=k_{\rm min}$, $k_2=k_{\rm min}$, then increasing $k_2$ and finally $k_1$. We find generally good agreement between the unwindowed and ideal bispectra, though there are some deviations on the largest scales due to the wide bins (spanning $k\in[0.05,0.45)\,\hMpc$ with $\Delta k = 0.05\,\hMpc$) adopted herein. The windowed bispectra show large residuals, whose size is quantified in the bottom panel (albeit for a small $0.4h^{-3}\mathrm{Gpc}^3$ survey).}
    \label{fig: bk-valid}
\end{figure}

\subsubsection{Simulation Results}
\noindent In Fig.\,\ref{fig: bk-valid}, we plot the bispectrum coefficients for simulations with and without an observational mask, using both the windowed and unwindowed estimators. As for the power spectrum, the mask induces strong deviations in the windowed bispectra, which are particularly notable on large-scales (with the first 20 bins containing the lowest $k$-modes). Here, the deviations are at the $2\sigma$ level compared to a $20\sigma$ overall detection (for this volume), and can potentially bias bispectrum inferences that derive constraining power from large scales, as shown in \citep{Chen:2024bdg}. In contrast, the unwindowed estimators perform significantly better, though we find slight biases in the lowest $k$ bins. This can occur since the mask alters the weighting of modes within a given bin, but can be greatly reduced if one adopts narrower bins at low-$k$ (where the spectrum varies significantly within a coarse bin).

\begin{figure}
    \centering
    \includegraphics[width=0.49\textwidth]{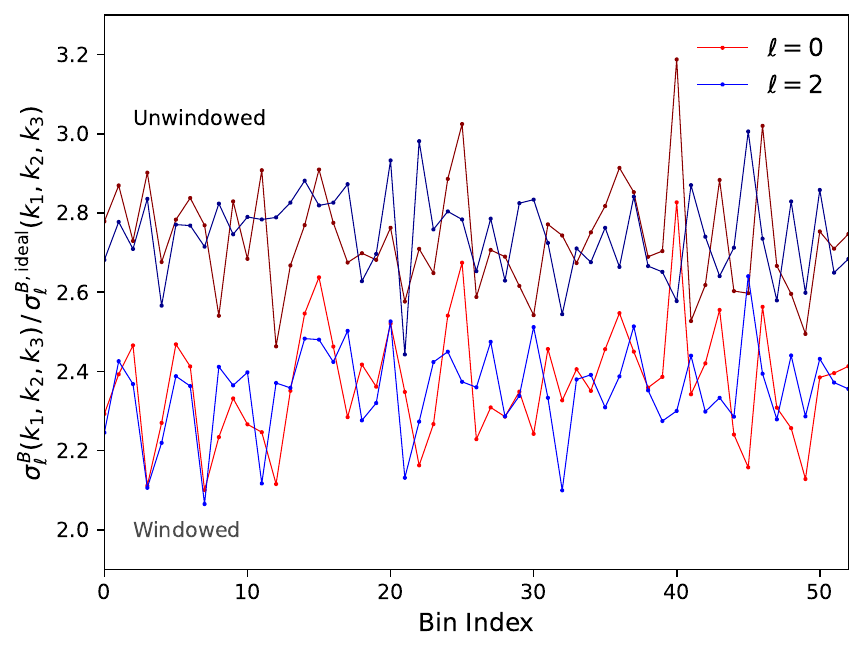}
    \includegraphics[width=0.49\textwidth]{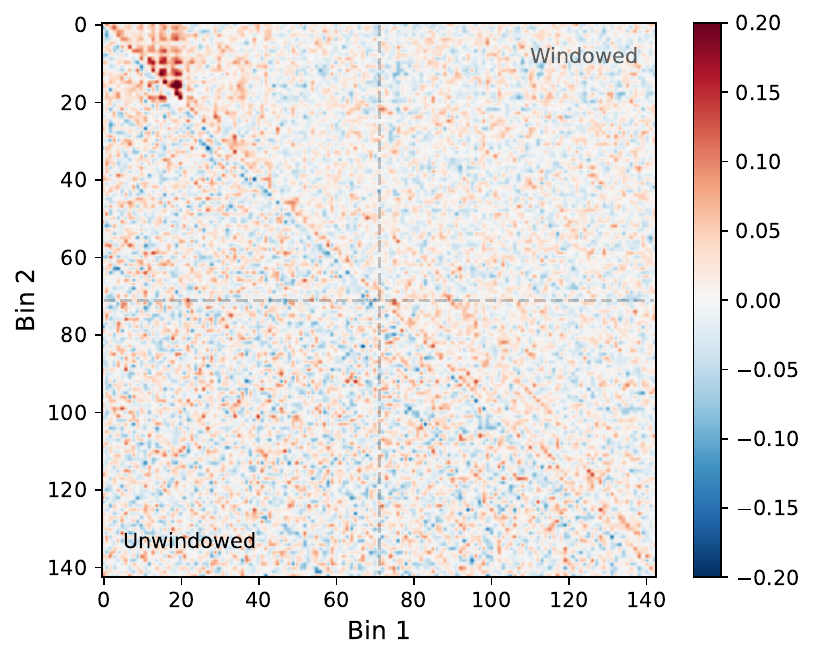}
    \caption{Comparison of the bispectrum covariances between unwindowed and windowed estimators, analogous to Fig.\,\ref{fig: pk-errors}. As for the power spectrum, the unwindowed measurements exhibit larger variances than their windowed equivalents (left), though this does not lead to loss of information due to the differing correlation structures (right). In particular, we find strong correlations between the first few bins in the window-convolved forms (upper left corner), which are not seen in the unwindowed correlations. We note that the off-diagonal correlations are, in general, enhanced if one uses narrower $k$-bins.}
    \label{fig: bk-errors}
\end{figure}

Fig.\,\ref{fig: bk-errors} compares the covariance of our two bispectrum estimators relative to that of ideal, unmasked, data (in all cases without an injected bispectrum). As for the power spectrum, we find larger variances for the unwindowed estimators than their windowed (and more conventional) equivalents; this is due to a difference in the correlation structure, and the unwindowed forms exhibit larger signal-to-noise over the bins of interest. In the windowed data, we find large covariances (at the $20\%$ level) between bins at low-$k$, particularly for squeezed configurations; since the underlying data is Gaussian, these are induced by the mask. In contrast, the unwindowed data shows much reduced bin-to-bin correlations, though the fine structure (which is itself damped by the wide $k$-bins) is shrouded by noise.

\subsubsection{Timing}
\noindent With the experimental parameters given above, the bispectrum numerators can be computed in \resub{$8$ seconds per simulation on a CPU node} (requiring 19 FFTs) \resub{or, when utilizing the \textsc{jax} backend, $\approx 1.2$ seconds on an A100 GPU}. This scales with the number of linear $k$-bins and the number of multipoles, given that we have assumed a global line-of-sight. In contrast, computation of the Fisher matrix (needed for the unwindowed estimators) scales with the total number of bins, \resub{and required $220$ seconds per iteration in our example ($600$ seconds on the GPU), with $556$ FFTs}. However, the increased variance of the bispectrum compared to the power spectrum allows \textit{far} fewer Monte Carlo iterations to be used to produce a converged bispectrum estimate: reducing $N_{\rm mc}$ from $500$ to $100$ induces an error at the $10^{-4}\sigma$ level, with only a $10^{-3}\sigma$ bias for $N_{\rm mc}=5$. This demonstrates the efficacy of the Girard-Hutchinson-type Monte Carlo summation methods, and demonstrates how our estimators can be very quickly computed in practice.

\begin{figure}
    \centering
    \includegraphics[width=0.6\textwidth]{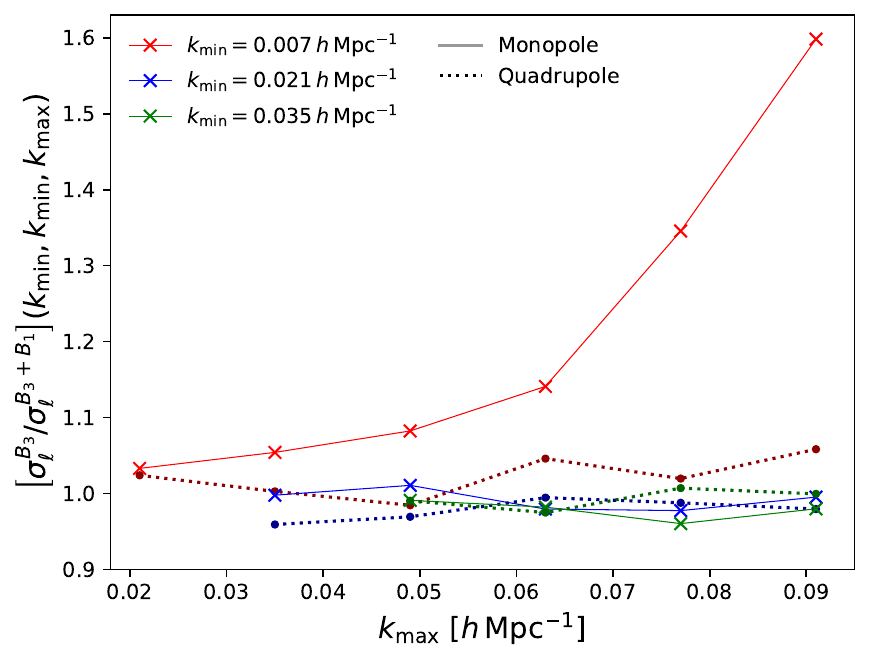}
    \caption{Impact of the linear term in the bispectrum estimators, as applied to masked data. We plot the ratio of error-bars obtained from $100$ bispectrum measurements computed excluding and including the linear term, which is itself estimated from $100$ Monte Carlo realizations. In all cases, data is masked using a BOSS-like window as in previous figures. Here, we focus on low $k$-modes, utilizing six linear bins with $k\in[0.007,0.091)\,\hMpc$, and plot a selection of isosceles bispectrum multipoles. We find significantly enhanced variances for squeezed triangles, demonstrating the utility of the linear term in constraining physics appearing in the squeezed limit.}
    \label{fig: bk-linear}
\end{figure}

\subsubsection{Linear Term}
\noindent Finally, we consider the formerly-neglected linear term of the bispectrum estimator. Motivated by the discussion in \S\ref{subsec: bk-ideal} (whence we note that the linear term vanishes in the ideal limit), we restrict our attention to large scales, using bins of width $\delta k = 0.014\,\hMpc$ with $k\in[0.007,0.091)\,\hMpc$. Bispectra are computed as before (using the FKP form for $\Si$), but we now include the linear term in the estimator, estimated from $100$ simulations (which is likely overly conservative). Whilst this significantly increases the runtime of the estimator (\resub{requiring $25$ seconds on a CPU-node or $75$ seconds on an A100 GPU, holding all maps in CPU memory}), Fig.\,\ref{fig: bk-linear} demonstrates that it gives significant improvements on the precision of squeezed bispectrum measurements, particularly when $k_{\rm min}$ approaches the fundamental frequency. Outside this regime, the gains are minimal, suggesting the utility of a hybrid approach to bispectrum estimation, whence the linear term is only used for squeezed configurations. This reduction in monopole error may significantly enhance the bounds on physics relevant on ultra-large scales, such as local-type primordial non-Gaussianity.

\section{Summary \& Conclusions}\label{sec: conclusion}
\noindent Robust estimation of correlation functions remains a central problem in cosmology. In the ideal translation-invariant limit, the optimal estimator is well-known \citep[e.g.,][]{2012PhRvD..86f3511F}; in observational settings, which usually feature spatially varying noise and mask, there is less consensus on what estimator should be used in practice. Many recent analyses of three-dimensional LSS data have utilized a simplified scheme, based on the FKP approximation, whence one weights the data by a local-in-configuration-space filter and computes the moments of the field directly \citep[e.g.,][]{1994ApJ...426...23F,2003ApJ...595..577Y,2006PASJ...58...93Y,2015MNRAS.453L..11B,2017JCAP...07..002H,2017JCAP...04..029S}. This has two drawbacks: (1) the weighting is optimal only in the limit of large-noise (at odds with future dense surveys); (2) the output (pseudo-)spectra are biased by the observational mask \citep[e.g.,][]{2004MNRAS.349..603E,Hamilton:2005kz}. Whilst (2) can be ameliorated by forward-modeling the effects of the mask on the theoretical $N$-point correlator (a $3(N-1)$-dimensional convolution integral), this procedure is computationally infeasible for $N>2$ (though see \citep{Pardede:2022udo} for recent advances), particularly when one needs to scan over many theoretical templates in a likelihood analysis. 

In this paper, we have considered general \textit{unwindowed} \resub{(and windowed)} estimators for the power spectrum and bispectrum of simulations and survey data, motivated by maximum-likelihood principles based on an Edgeworth expansion \citep[e.g.,][]{1998ApJ...499..555T,Hamilton:2005kz,Hamilton:2005ma,Hamilton:1999uw} (see \citep{Philcox:2023uwe,Philcox:2023psd,Philcox4pt3,Philcox:2025lxt} for recent CMB applications). Due to the particular choice of our normalization matrix, $\F$, these estimators are not biased by the mask on average (hence, `unwindowed'), and can also be applied to arbitrarily weighted data. The latter point is particularly relevant in observational contexts, when one may wish to apply Wiener filtering, mode deprojection, inpainting, or a simple FKP-like weight, and additionally facilitates computation of the \textit{optimal} estimators (whose weight satisfies an $N_{\rm pix}\times N_{\rm pix}$ matrix equation). We additionally place close attention to holes in the observational mask (\textit{i.e.} a non-invertible mask), which can cause numerical instabilities and bias if not correctly accounted for. The result of this study in linear algebra is a set of (optionally optimal) estimators for the anisotropic power spectrum and bispectrum, that, employing various computational tricks, can be entirely formulated in terms of Fourier transforms and Monte Carlo summation. These are efficient to compute (scaling at most as $\mathcal{O}(N_{\rm bins}N_{\rm pix}\log N_{\rm pix})$) and return unbiased and minimum-variance estimates of the underlying spectra.

Accompanying this work is a new code package, \polybin, which implements the above estimators in \textsc{Python}\resub{/\textsc{Cython}}, as well as their idealized limits, which are intended to be used for the study of $N$-body simulations \resub{or for computing standard windowed power spectra from FKP-weighted data}. The code additionally provides support for GPU acceleration using \textsc{jax}, which can significantly reduce execution times. We provide an extensive set of tutorials describing the functionality of \polybin, and showing a number of practical use-cases. Furthermore, we have presented an extensive set of validation tests for both the power spectrum and bispectrum estimators, with the following conclusions: \resub{(a) our estimators agree with previous codes to excellent precision, where relevant}, (b) regardless of the weighting scheme, the estimators are not biased by the survey geometry (assuming sufficiently thin bins), (c) optimal weights (and thus minimum-variance errors) can be practically implemented via CGD methods, and yield slight ($5-10\%$) improvements on the power spectrum errorbars on large-scales, (d) the inverse Fisher matrix $\F^{-1}$ provides a useful proxy for the (masked) Gaussian power spectrum covariance if the weights are optimal, (e) the inclusion of a linear term in the bispectrum estimator can significantly reduce noise when one analyses squeezed bispectra (e.g., for local primordial non-Gaussianity), (f) \resub{one can compute Gaussian covariance matrices for masked fields analogously to the power spectrum normalization}.

Finally, we consider the drawbacks of the above approaches, as well as their extensions. Due to the need to compute an $N_{\rm bins}\times N_{\rm bins}$ coupling matrix for each statistic, the unwindowed polyspectrum estimates are naturally more expensive to compute than their simplified ``pseudo-spectrum'' equivalents; \resub{however, this increase is somewhat artificial, since we no longer need to compute window functions explicitly (indeed, the normalization matrices \textit{are} window functions)}. Whilst we have here demonstrated how such costs can be substantially mitigated by utilizing Monte Carlo methods (ensuring scalings linear in $N_{\rm bins}$ and using only FFTs) and by noting that the matrix can be computed independently from the data, this nevertheless represents an important limitation of the approach. This is particularly apparent if one attempts to compute \resub{bispectra} with limited computational memory, whence construction of $\F$ as an outer product becomes infeasible. We note however, that such a difficulty appears also in the more common approach of forward-modeling pseudo-spectra; there, one must either compute the forward-modeling matrix (which can be prohibitively expensive), or make simplifying assumptions (which could induce bias). For modeling higher-point functions, we expect that the data-oriented ``unwindowing'' approach of this work could be more efficient, since one does not rely on computing mask-distortions theoretically (which is expensive, and, at heart, still requires counting random points). In this vein, it would be interesting to extend the algorithms of this work to correlators beyond the bispectrum. This would facilitate a wide range of analyses, such as a first measurement of cubic primordial non-Gaussianity in large-scale structure.

\acknowledgments
{\footnotesize
\begingroup
\hypersetup{hidelinks}
\noindent 
We thank Chirag Modi for insightful discussions on stochastic trace estimation, as well as Emiliano Sefusatti for motivating this work. \resub{We are additionally grateful to the anonymous referee for insightful comments}. OHEP thanks Bagels \& Co for Sunday sustenance. OHEP is a Junior Fellow of the Simons Society of Fellows. TF is supported by the Fundamentals of the Universe research program at the University of Groningen, and thanks the Center for Information Technology of the University of Groningen for providing access to the Hábrók high-performance computing cluster.
\endgroup
}

\appendix
\section{Theoretical Binning Matrices}\label{app: binning-theory}
\noindent In this appendix, we consider how to compute theoretical predictions for the $N_{\rm bins}$ observed bandpowers $\hat{p}_\alpha$, given some finely-binned power spectrum model $p^{\rm fine}_\iota$ where $\iota\in\{1,2,\cdots,N_{\rm fine}\}$. If the data and theory bins align, this is straightforward: the expected value of $\hat{p}_\alpha$ is simply $p^{\rm fine}_\alpha$, as expected. In essence, this approach forward models the effect of bin-convolution on the output statistic, $\hat{p}_\alpha$ (retaining the $\F_{\alpha\beta}^{-1}$ factor, to ensure \textit{approximate} mask-deconvolution, \resub{analogous to the pseudo-$C_\ell$ technique used in CMB analyses} \citep{Alonso:2018jzx}).

Starting from \eqref{eq: pk-estimator-def}, we can compute the expectation of $p_\alpha^{\rm num}$ by inserting the finely-binned definition $\xi = \sum_{\iota=1}^{N_{\rm fine}} p_{\iota}^{\rm fine}\,\partial\xi/\partial p^{\rm fine}_\iota$ into the expectation $\av{dd^\dag}$. This yields
\beq\label{eq: theory-matrix-formal}
    \av{p^{\rm num}_\alpha}&=&\sum_{\iota=1}^{N_{\rm fine}}\G_{\alpha\iota}p^{\rm fine}_\iota, \qquad 
    \G_{\alpha\iota} = \frac{1}{2}\mathrm{Tr}\left(\frac{\partial\xi}{\partial p_\alpha}\cdot[\Si \P]\cdot\frac{\partial\xi}{\partial p^{\rm fine}_\iota}\cdot[\Si \P]^\dag\right),
\eeq
in terms of the $N_{\rm bins}\times N_{\rm fine}$ matrix $\G_{\alpha\iota}$, which is a rectangular analogue of the usual Fisher matrix. In full, the theory predictions are given by $\av{\hat{p}_\alpha} = \sum_{\iota}\left(\sum_\beta\F^{-1}_{\alpha\beta}\G_{\beta\iota}\right)p^{\rm fine}_\iota$; if the bins are suitably thin, this is well approximated by the theory spectrum at the bin-centers, reproducing the previous results. 

As in \S\ref{subsec: pk-implementation}, the trace term in \eqref{eq: theory-matrix-formal} can be rewritten as an average over random fields $a$ with covariance $\A$ (cf.\,\ref{eq: fish-from-Q}): 
\beq\label{eq: theory-matrix-def}
    \resub{\G_{\alpha\iota} = \frac{1}{2}\av{\left(\P^\dag\Sit\mathsf{Q}_\alpha[\Si \P a]\right)^\dag\cdot\mathsf{Q}^{\rm fine}_\iota[\Ai a]}_a},
\eeq
defining $\mathsf{Q}^{\rm fine}_{\iota}[u] = \partial\xi/\partial p_\iota^{\rm fine}\cdot u$. As noted previously, we can drop the symmetrization in $\mathsf{Q}_\iota^{\rm fine}$; this implies that the number of FFTs used to compute $\G_{\alpha\iota}$ (and thus the dominant computational cost) is independent of $N_{\rm fine}$, scaling only with $N_{\rm bins}$. The above algorithm is implemented within \polybin (in the \texttt{compute\_theory\_contribution} code), and validated in the tutorial scripts found online. We do not attempt to compute an analogue for the bispectrum and beyond, since this is highly computationally expensive, and one cannot adopt the same tricks as above to remove dependence on $N_{\rm fine}$.

\bibliographystyle{apsrev4-1}
\bibliography{refs}

\end{document}